\documentclass[aps,pra,superscriptaddress,twocolumn,longbibliography,floatfix]{revtex4-2}
\usepackage{units}
\usepackage{amsmath}
\usepackage{amsthm}
\usepackage{amssymb}
\usepackage{graphicx}
\usepackage[english]{babel}
\usepackage[utf8]{inputenc}
\usepackage{mathtools}
\usepackage{color}
\usepackage{xcolor}
\usepackage{bbold}
\usepackage{braket}
\usepackage{tabularray}
\usepackage[version=3]{mhchem}
\usepackage{svg}

\definecolor{myurlcolor}{rgb}{0,0,0.7}
\definecolor{myrefcolor}{rgb}{0.8,0,0}

\usepackage[
	breaklinks,
	pdftex,
	colorlinks=true, 
	linkcolor=myrefcolor,
	citecolor=myurlcolor,
	urlcolor=myurlcolor
]{hyperref}
\usepackage{cleveref}

\usepackage[linesnumbered, ruled,vlined]{algorithm2e}

\graphicspath{{./images/},{./imagesAppendix/},{./Plots/}}

\renewcommand{\eqref}[1]{Eq.~(\ref{#1})} 

\def\app#1#2{%
  \mathrel{%
    \setbox0=\hbox{$#1\sim$}%
    \setbox2=\hbox{%
      \rlap{\hbox{$#1\propto$}}%
      \lower1.1\ht0\box0%
    }%
    \raise0.25\ht2\box2%
  }%
}

\ifx\proof\undefined
\newenvironment{proof}[1][\protect\proofname]{\par
	\normalfont\topsep6\p@\@plus6\p@\relax
	\trivlist
	\itemindent\parindent
	\item[\hskip\labelsep\scshape #1]\ignorespaces
}{%
	\endtrivlist\@endpefalse
}
\providecommand{\proofname}{Proof}
\fi
\makeatother

\providecommand{\ftname}{ft}
\providecommand{\theoremname}{Theorem}
\providecommand{\claimname}{Claim}
\providecommand{\lemmaname}{Lemma}
\providecommand{\definitionname}{Definition}

\definecolor{KB}{rgb}{0.4,0.3,0.9}

\definecolor{THc}{rgb}{0.9,0.3,0.2}

\newcommand{\sectionMain}[1]{
\let\oldaddcontentsline\addcontentsline
\renewcommand{\addcontentsline}[3]{}
\section{#1}
\let\addcontentsline\oldaddcontentsline
}

\allowdisplaybreaks

\begin{document}

\title{Controlled dissipation for Rydberg atom experiments}

\author{Bleuenn B\'egoc}
\affiliation{CNR-INO, via G. Moruzzi 1, 56124 Pisa, Italy}
\affiliation{Dipartimento di Fisica dell’Universit\`{a} di Pisa, Largo Pontecorvo 3, 56127 Pisa, Italy}

\author{Giovanni Cichelli}
\thanks{Present Address: Dipartimento di Fisica, Politecnico di Milano, Piazza Leonardo da Vinci 32, 20133 Milano, Italy}
\affiliation{Dipartimento di Fisica dell’Universit\`{a} di Pisa, Largo Pontecorvo 3, 56127 Pisa, Italy}

\author{Sukhjit P. Singh}
\thanks{Present Address: Centre for Quantum Dynamics and Centre for Quantum Computation and Communication Technology, Griffith University, Yuggera Country, Brisbane, QLD, 4111, Australia}
\affiliation{Dipartimento di Fisica dell’Universit\`{a} di Pisa, Largo Pontecorvo 3, 56127 Pisa, Italy}

\author{Fabio Bensch}
\thanks{Present Address: Physikalisches Institut, Universit\"at T\"ubingen, Auf der Morgenstelle 14, 72076 T\"ubingen, Germany}
\affiliation{Dipartimento di Fisica dell’Universit\`{a} di Pisa, Largo Pontecorvo 3, 56127 Pisa, Italy}

\author{Valerio Amico}
\thanks{Present Address: Physikalisches Institut, Universit\"at T\"ubingen, Auf der Morgenstelle 14, 72076 T\"ubingen, Germany}
\affiliation{Dipartimento di Fisica dell’Universit\`{a} di Pisa, Largo Pontecorvo 3, 56127 Pisa, Italy}

\author{Francesco Perciavalle}
\address{Quantum Research Center, Technology Innovation Institute, P.O. Box 9639 Abu Dhabi, UAE}
\address{Dipartimento di Fisica dell’Universit\`a di Pisa and INFN, Largo Pontecorvo 3, I-56127 Pisa, Italy}

\author{Davide Rossini}
\address{Dipartimento di Fisica dell’Universit\`a di Pisa and INFN, Largo Pontecorvo 3, I-56127 Pisa, Italy}

\author{Luigi Amico}
\affiliation{Quantum Research Center, Technology Innovation Institute, P.O. Box 9639 Abu Dhabi, UAE}
\affiliation{Dipartimento di Fisica e Astronomia ``Ettore Majorana'', Via S. Sofia 64, 95123 Catania, Italy}
\affiliation{INFN-Sezione di Catania, Via S. Sofia 64, 95123 Catania, Italy}

\author{Oliver Morsch}
\affiliation{CNR-INO, via G. Moruzzi 1, 56124 Pisa, Italy}
\affiliation{Dipartimento di Fisica dell’Universit\`{a} di Pisa, Largo Pontecorvo 3, 56127 Pisa, Italy}

\begin{abstract}
    We demonstrate a simple technique for adding controlled dissipation to Rydberg atom experiments. In our experiments we excite cold rubidium atoms in a magneto-optical trap to $70$-S Rydberg states, whilst simultaneously inducing forced dissipation by resonantly coupling the Rydberg state to a hyperfine level of the short-lived $6$-P state. The resulting effective dissipation can be varied in strength and switched on and off during a single experimental cycle. 
\end{abstract}

\maketitle

\section{Introduction}
In recent years, ultra-cold Rydberg atoms have been extensively studied, in particular in the context of quantum computation and simulation~\cite{Saffman2016, Browaeys2020132,  Morgado2021}. In those applications, dissipation and decoherence are usually detrimental and need to be minimized. However, there are also contexts in which dissipation is a feature, such as in the study of driven-dissipative systems exhibiting nonequilibrium phase transitions~\cite{Gutierrez2017, Morsch2018383, Helmrich2020481, Brennecke2013, Rodriguez2017, Fink2017, Fink2017-2, Fitzpatrick2017, Cai2021, LiZeno2018, PhysRevB.105.L241114}, or in the dissipative preparation of entangled states~\cite{Diehl2008, Verstraete2009, Muller20121, Rao2013, Carr2013, Rao2014, Lee2015, Su2015,  Shao2017, Roghani2018, Harrington2022660, Yang2021}.
\begin{figure}[!t]
    \centering
\includegraphics[width=0.5\textwidth]{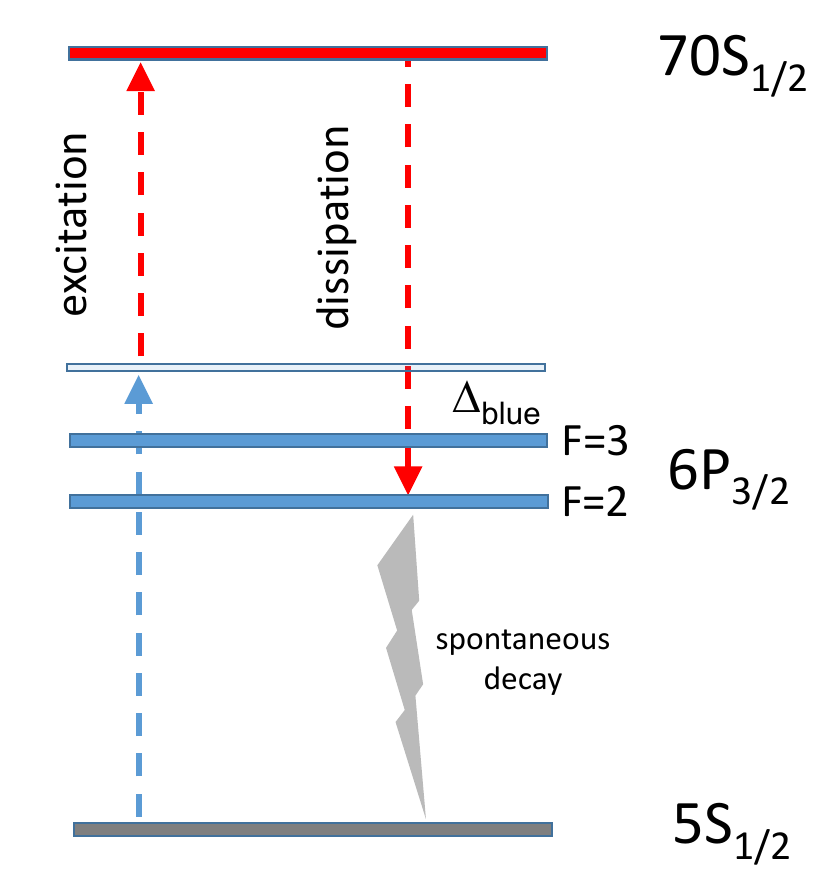}
    \caption{Scheme for simultaneous excitation and controlled dissipation in rubidium Rydberg atoms. The 6P level is used both as an intermediate state for two-photon excitation (blue and red arrows on the left) and for resonant depumping from the 70S Rydberg state (red arrow on the right). The energy difference between the two 6P hyperfine levels used here is $h\times86.97 \,\mathrm{MHz}$.}
    \label{Fig1}
\end{figure}
Dissipation in Rydberg atom experiments arises naturally from spontaneous emission and black-body radiation induced transitions between Rydberg levels~\cite{Beterov2009,Archimi2022}. The black-body radiation induced dissipation can lead to a substantial departure from an approximate two-level dynamics (either ground state and Rydberg state or two Rydberg levels), making the interpretation of experiments difficult.

Here we show results on a simple method than implements controllable dissipation in a Rydberg system. We study the dynamics of the system under simultaneous excitation and controlled dissipation and find that the number of excitations tends towards a steady state determined by the ratio of the total excitation and dissipation rates. Our technique can be used to control the dissipative timescale in experiments on driven-dissipative systems and also pave the way towards the dissipative generation of correlated many-body quantum states of Rydberg ensembles.

\section{Experimental setup and results}

\begin{figure}[!h]
    \centering
\includegraphics[width=0.5\textwidth]{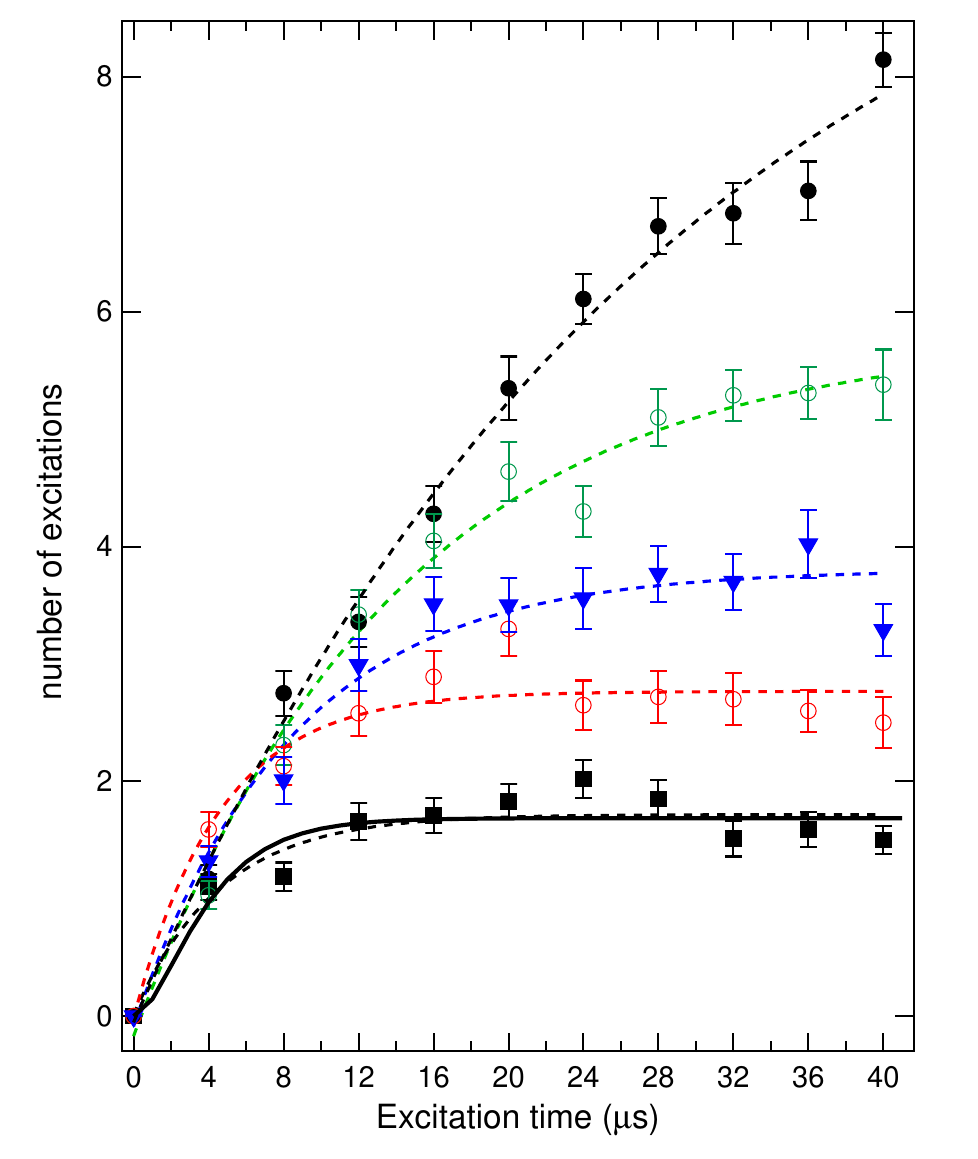}
    \caption{Simultaneous excitation and controlled dissipation of Rydberg atoms. The two-photon excitation Rabi frequency is $2\pi\times 12 \, \mathrm{kHz}$, corresponding to a single-particle excitation rate of $0.65\,\mathrm{kHz}$, and the Rabi frequencies of the dissipation beam are: $0$ (black circles), $2\pi\times 0.25\,\mathrm{MHz}$ (open green circles), $2\pi\times 0.47\,\mathrm{MHz}$ (blue triangles), $2\pi\times 0.64\,\mathrm{MHz}$ (open red circles) and $2\pi\times 1.1\,\mathrm{MHz}$ (black squares). Dashed lines are fits with exponential functions to guide the eye. The solid line is a numerical simulation of a three-level system with a Rabi frequency of the dissipation beam $\Omega_{\rm diss}=2\pi\times 0.239\,\mathrm{MHz}$, a Rabi frequency of the excitation beam $\Omega_{\rm exc}=2\pi\times 16\,\mathrm{kHz}$, and a reference value of the intermediate state decay rate of $7$MHz. We take into account the presence of pure dephasing with rate $1$MHz on the ground-Rydberg transition; for details on the simulation see the Appendix. Error bars are one standard error (averaged over $100$ repetitions).
    }
    \label{Fig2}
    
\end{figure}
In our experiments, we start from an ultra-cold cloud of 87-Rb atoms in a magneto-optical trap (MOT). The atomic cloud has a roughly Gaussian shape in the three spatial directions of width around $100\,\mathrm{\mu m}$, with a peak density of a few times $10^9\,\mathrm{cm^{-3}}$. To induce excitation to the $70S$ Rydberg state from the $5S$ ground state, we use a two-photon process via the intermediate $6P$ level with laser beams at $420\,\mathrm{nm}$ (blue) and $1013\,\mathrm{nm}$ (IR) (Gaussian shape, waists $40\,\mathrm{\mu m}$ and $90\,\mathrm{\mu m}$, respectively), where the $420\,\mathrm{nm}$ laser is detuned by $\Delta_\mathrm{blue}=70\,\mathrm{MHz}$ with respect to the $6P$, $F=3$ hyperfine level (see Fig.~\ref{Fig1}). 

Controlled dissipation is induced by another laser beam at $1013\,\mathrm{nm}$ (Gaussian shape, waist $250\,\mathrm{\mu m}$, power up to $30\,\mathrm{mW}$) which is resonant with the transition $6P, F=2 \longrightarrow 70S$. Both 1013 nm laser beams are derived from the same laser by splitting the output of the laser and shifting the wavelengths of the two resulting beams using acousto-optic modulators (AOMs), which are also used to control the powers of the two beams. Rydberg atoms are then detected via field ionization and ion detection using a channeltron~\cite{Simonelli2017}, with an overall detection efficiency of around $40$ percent (in this paper, we report the detected number of excitations without correcting for the detection efficiency). 

Figure~\ref{Fig2} shows the general principle used for the controlled dissipation. As shown in Ref.~\cite{Simonelli2017}, an effective depumping from the Rydberg state can be achieved by resonantly coupling the Rydberg state to the fast decaying $6P$ level (lifetime $\tau_{6P}=118\,\mathrm{ns}$). In this way, we can induce dissipation rates up to $250\,\mathrm{kHz}$, compared to around $5\,\mathrm{kHz}$ for spontaneous decay from the $70S$ state. The intermediate state decays sufficiently fast so that its population is approximately $0$ during the entire process.

In Fig.~\ref{Fig2} we show the results of a typical experiment in which atoms are excited to the 70S state whilst the dissipation beam is constantly depumping the atoms at varying rates. For the largest depumping rates, the number of excitations reaches a steady state after around $10$ microseconds. Even without the controlled dissipation, the time-dependence of the number of excitations is not linear, because of dipole blockade effects~\cite{Comparat2010A208,Valado2016} and the finite natural decay rate of $5\,\mathrm{kHz}$. From the experimentally measured total excitation rate $\Gamma_{\rm exc}^{(\rm tot)}$(obtained from the initial slope of the curve without controlled dissipation) of around $500\,\mathrm{kHz}$ and the steady-state number of excitations $N_{\rm ss}$, we used the relationship $N_{\rm ss}=\Gamma_{\rm exc}^{(\rm tot)}/\Gamma_{\rm diss}$ infer a maximum controlled dissipation rate $\Gamma_{\rm diss}$ of around $250\,\mathrm{kHz}$.

\begin{figure}[!t]
    \centering
\includegraphics[width=0.5\textwidth]{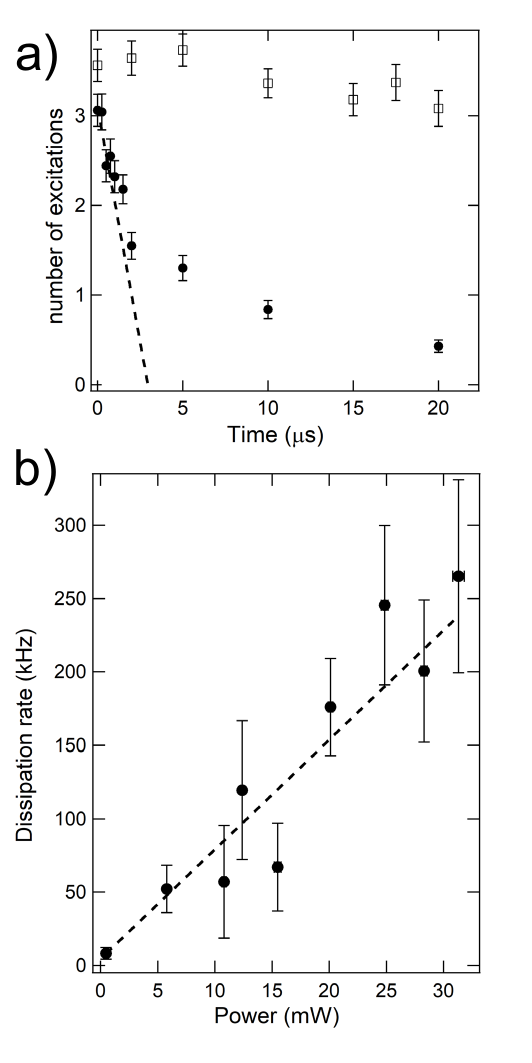}
    \caption{Measurement of the controlled dissipation rate. (a) Number of excitations as a function of time with only the dissipation laser switched on, after a brief initial excitation pulse. The dissipation rate is measured by performing a linear fit (dashed line) to the (approximately) linear part of the decay curve, shown here for $28\,\mathrm{mW}$ power of the dissipation laser (filled circles). Also shown is the decay curve without the dissipation laser (open squares). (b) Measured dissipation rates as a function of the power of the dissipation laser. The dashed line is a linear fit to guide the eye. 
    }
    \label{Fig3}
\end{figure}

We have also measured the rate of controlled dissipation $\Gamma_{\rm diss}$ directly, by first creating a small number of Rydberg excitations using a short excitation pulse (around $5\,\mathrm{\mu s}$). Thereafter, the atomic cloud was exposed only to the dissipation laser for a variable time, after which the number of excitations was measured. The dissipation rate was then calculated by fitting a straight line to the initial, approximately linear part of the decay curve, and dividing the slope obtained by the initial number of excitations. The results are shown in Fig.~\ref{Fig3}, showing the expected linear dependence of the dissipation rate on the power of the dissipation laser, and confirming a maximum dissipation rate of $250\,\mathrm{kHz}$.

To compare our experimental results to theoretical expectations based on the measured laser parameters, we calculate the averages of the two-photon excitation Rabi frequency $\Omega_{\rm exc}=\Omega_{\rm IR}\Omega_{\rm blue}/(2\Delta_{\rm blue})$ and the dissipation Rabi frequency $\Omega_{\rm diss}$ over a volume defined by the widths of the respective lasers involved, using the ARC software package~\cite{Sib2017}. For the maximum power in the dissipation beam, we obtain a theoretical $\Omega_{\rm diss}$ of around $2 \pi \times  1\,\mathrm{MHz}$, from which we estimate a maximum dissipation rate (see Appendix for the derivation) $\Gamma_{\rm diss}^{(\rm theo)}=\Omega_{\rm diss}^2/(\gamma_{6P})\approx 4.6\,\mathrm{MHz}$ (here, $\gamma_{6P}=1/\tau_{6P}$ is the linewidth of the $6P$ intermediate state). This is roughly an order of magnitude larger than the maximum dissipation rate we measure experimentally. While we currently have no simple explanation for this discrepancy, we believe that it is a combination of imperfections in the beam alignment and in the intensity profile (deviation from a perfect Gaussian) of the dissipation beam. 

\begin{figure}[!h]
    \centering
\includegraphics[width=0.5\textwidth]{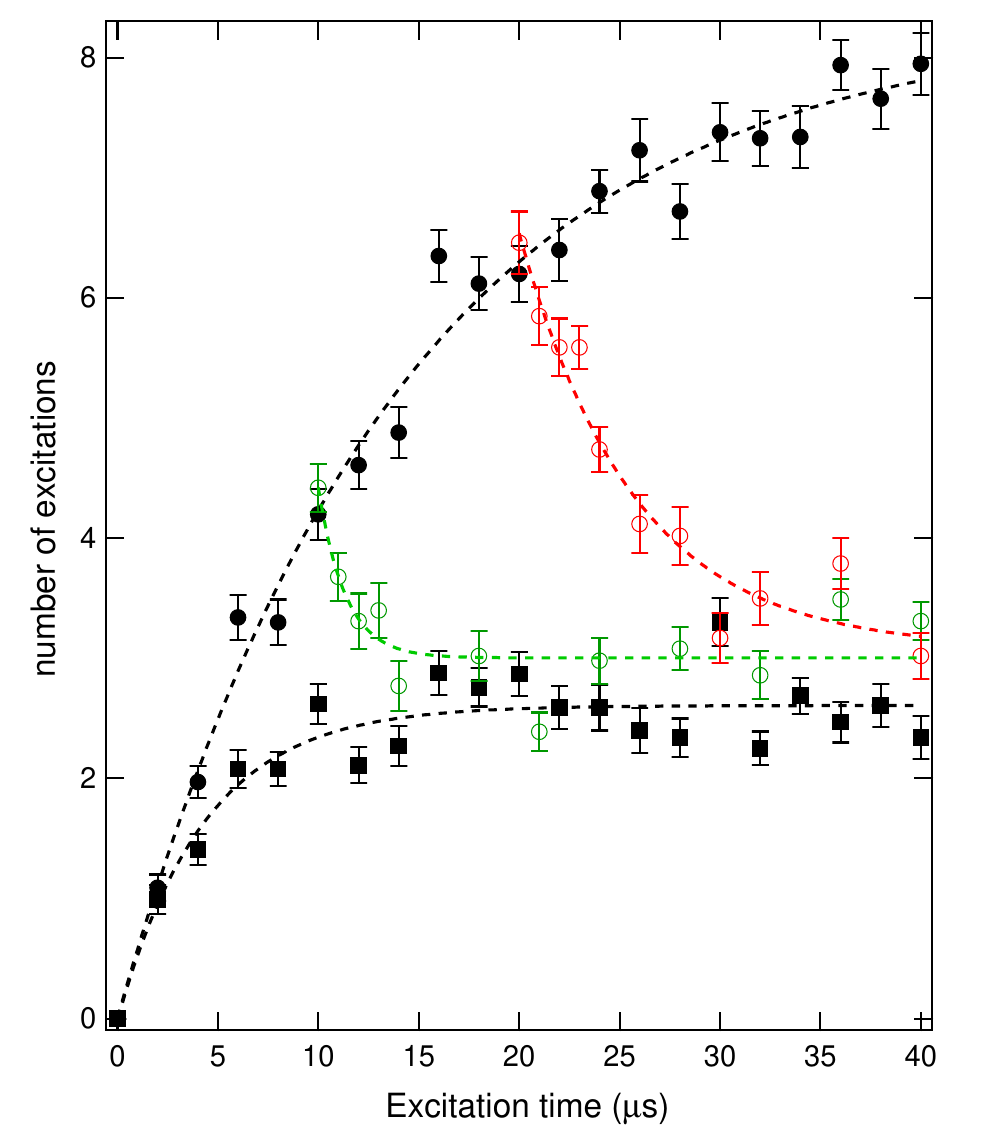}
    \caption{Time-controlled dissipation. The dissipation laser is switched on at $t=0$ (black squares), $t=10\,\mathrm{\mu s}$, and $t=20\,\mathrm{\mu s}$ (black circles: no controlled dissipation). Dashed lines are fits with exponential functions to guide the eye. The excitation Rabi frequency is $2\pi\times 9.6 \,\mathrm{kHz}$ and the dissipation rate is $250\,\mathrm{kHz}$.
    }
    \label{Fig4}
\end{figure}

We now demonstrate time-control of the induced dissipation. In Fig.~\ref{Fig4} we show results of an experiment in which the controlled dissipation (measured rate around $85\,\mathrm{kHz}$) was switched on at different times during an experimental cycle. For switch-on times $0\,\mathrm{\mu s}$, $10\,\mathrm{\mu s}$, and $20\,\mathrm{\mu s}$, the number of excitations tends to roughly the same steady-state value, as expected. For non-zero switch-on times, the steady-state value appears to be slightly larger than for switch-on at $t=0$, This might be due to interaction effects, since the dissipation rate for each atom will depend on the presence of nearby Rydberg atoms, which can shift the Rydberg level through the van der Waals interaction, similarly to the dipole blockade effect~\cite{Comparat2010A208,Valado2016}.

\section{Conclusions}
We have demonstrated controlled dissipation in a Rydberg atom experiment. We have shown that the dissipation rate can be easily controlled in time over the course of a single experiment. Our scheme can be straightforwardly extended to spatially varying dissipation, for instance using a digital micro-mirror device to design appropriately shaped intensity profiles of the dissipation beam. 

Since, in the present experiment, we are using a narrow-linewidth ($<1\,\mathrm{MHz}$) laser beam to induce dissipation, the dissipation rate of an individual Rydberg atom will depend on the state of nearby atoms, as any nearby atoms in Rydberg states will induce a level shift and hence a variation in the dissipation rate, in analogy with the dipole blockade (if the dissipation laser is resonant) and facilitation (if the dissipation laser is off-resonant). This can, in principle, be avoided by using a larger linewidth, or artificially broadened, laser (tens of MHz) at the expense of a proportionally larger total power required to achieve the same dissipation rates. On the other hand, the blockade and facilitation effects on dissipation could also be exploited in order to study models with correlated dissipation~\cite{Nalbach2015,Duan19974466,Masson2022,Chakrabarti2023,kazemi2023driven}. For instance, by tuning the  dissipation laser off resonance at a specific time,  only those Rydberg atoms that have a neighbouring excited atom at the facilitation distance  will be subject to controlled dissipation. Such a scheme may be adopted for conceiving new devices based on the control of facilitation in Rydberg atoms~\cite{kitson2023rydberg}.

\acknowledgments
We thank Frederico Brito for useful discussions. The Julian Schwinger Foundation grant JSF-18-12-0011 is acknowledged. OM and BB also acknowledge support by the H2020 ITN ``MOQS" (grant agreement number 955479), and OM acknowledges MUR (Ministero dell’Università e della Ricerca) through the PNRR MUR project PE0000023-NQSTI.

\appendix

\section{Effective two-level description}
\begin{figure}[!h]
    \centering
\includegraphics[width=0.75\columnwidth]{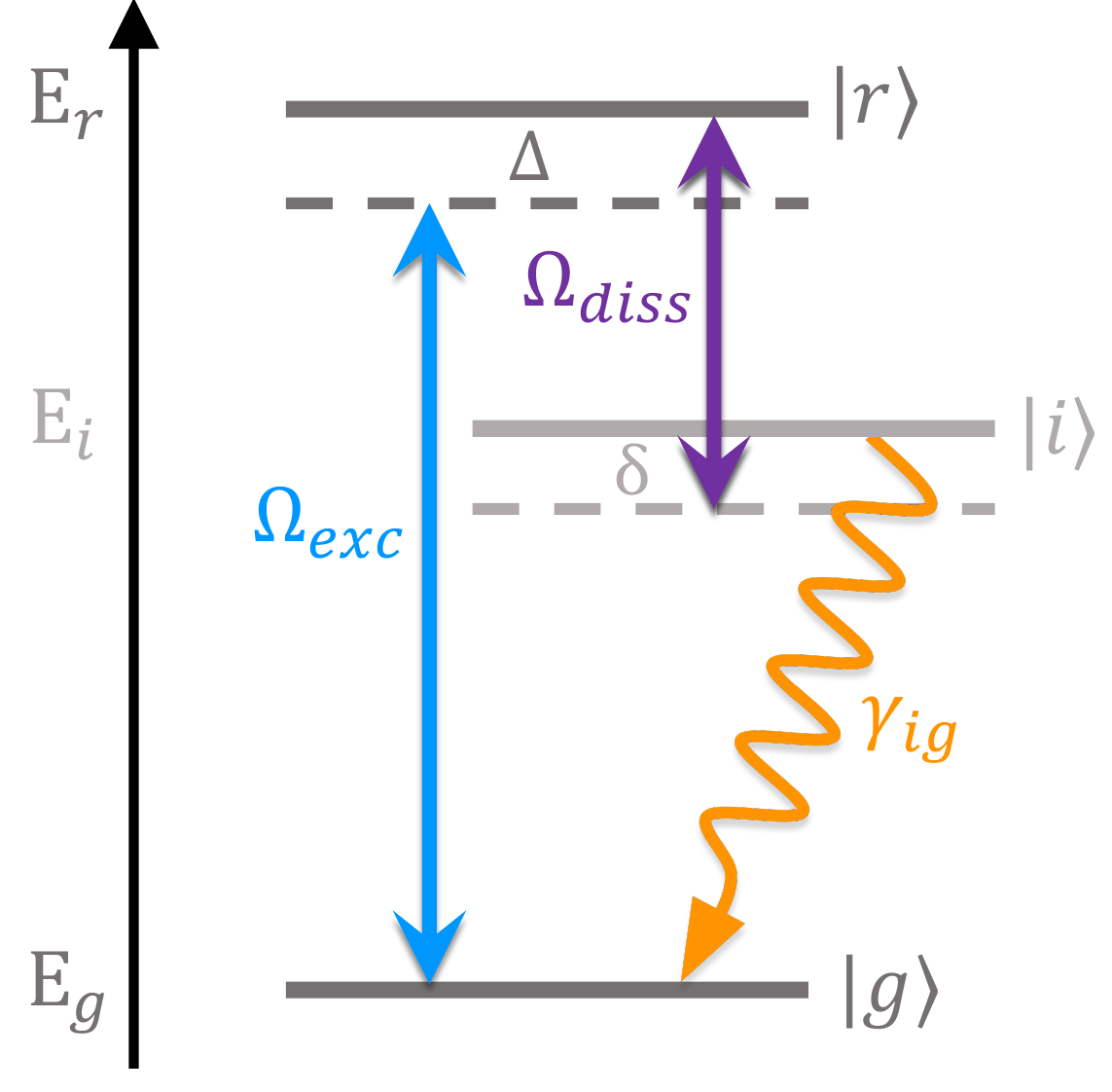}
    \caption{Three-level system composed of the atomic ground state $\ket{g}$, an highly excited Rydberg state $\ket{r}$, and an intermediate state $\ket{i}$. The energies of the three states are $E_g$, $E_i$, and $E_r$, respectively. The Rydberg and ground states are coupled through the excitation field $\Omega_{\rm exc}$, while the Rydberg and intermediate states through the dissipation laser $\Omega_{\rm diss}$. Here $\gamma_{ig}$ is the decay rate of the intermediate state, while $\Delta$ and $\delta$ are the detunings of the excitation and dissipation fields, with respect to the Rydberg and the intermediate state.}
    \label{app:sketch}
\end{figure}
To qualitatively describe the problem, we employ a single-atom description. The original model is composed by a three-level system composed by ground, Rydberg, and intermediate states, where a Rydberg-intermediate coupling together with its finite lifetime is used to engineer the dissipation. Our setup can be mapped into an effective two-level system, in which the Hamiltonian parameters can be used to control the dissipation, in particular, the decay of the Rydberg state into the ground state. To show this, we rely on the adiabatic elimination approach~\cite{ReiterEffective2012, XaoHigh2023, Schempp2015transport, kazemi2023driven}.

We consider a three-level system composed of the ground state $\ket{g}$, an intermediate state $\ket{i}$, and the Rydberg state $\ket{r}$, such that $E_g < E_i < E_r$. We assume that the decay rate of $\ket{r}$ is negligible, while $\ket{i}$ has a finite lifetime $\gamma_{ig}^{-1}$. We couple $\ket{g}$ and $\ket{r}$ through an excitation field with frequency $\omega_{\rm exc}$ and Rabi frequency $\Omega_{\rm exc}$. To engineer the decay of the Rydberg state $\ket{r}$ into the ground state $\ket{g}$, we couple $\ket{r}$ with the intermediate state $\ket{i}$ through a field with frequency $\omega_{\rm diss}$ and Rabi frequency $\Omega_{\rm diss}$. In the main text we considered this type of scheme in which the ground state is $\ket{5S_{1/2}}$, the intermediate state $\ket{6P_{3/2}}$, and the Rydberg state is $\ket{70S_{1/2}}$. 

The general Hamiltonian describing such system in the rotating wave approximation is
\begin{eqnarray}
\dfrac{\hat{H}(t)}{\hbar} & = & \omega_i \! \ket{i}\!\bra{i} + \omega_r \! \ket{r}\!\bra{r}+\dfrac{\Omega_{\rm diss}}{2} \left(e^{-i\omega_{\rm diss} t}\ket{r}\!\bra{i} + {\rm H.c.}\right) \nonumber \\&& + \! \dfrac{\Omega_{\rm exc}}{2} \! \left(e^{-i\omega_{\rm exc} t} \! \ket{r}\!\bra{g} + {\rm H.c.}\right),
\label{eq:H_time}
\end{eqnarray}
where $\omega_i=E_i/\hbar$ and $\omega_r=E_r/\hbar$ are the angular frequencies associated with the intermediate and Rydberg states respectively, while the ground-state energy $E_g$ has been set to zero. We rotate out the time dependence of the Hamiltonian using the unitary transformation $\hat{\mathcal{U}}(t)=\ket{g}\bra{g}+e^{i(\omega_{\rm exc} - \omega_{\rm diss})t}\ket{i}\bra{i} + e^{i\omega_{\rm exc}t}\ket{r}\bra{r}$ and we obtain
\begin{eqnarray}
\label{eq:H_rot}
\dfrac{\hat{H}}{\hbar} & = & (\delta + \Delta)\ket{i}\bra{i}+\Delta\ket{r}\bra{r} \\ 
& & +\dfrac{\Omega_{\rm diss}}{2} \left(\ket{r}\bra{i} +  {\rm H.c.} \right) + \dfrac{\Omega_{\rm exc}}{2} \left( \ket{r}\bra{g} + {\rm H.c.} \right), \nonumber  
\end{eqnarray}
where we defined the detunings $\Delta=\omega_r-\omega_{\rm exc}$ and $\delta=\omega_{\rm diss}-(\omega_r - \omega_i)$.
We include the intermediate state decay through a decay term in the Lindblad master equation~\cite{breuer2002theory}
\begin{equation}
    \dfrac{\partial \rho}{\partial t}=-i \bigg[ \frac{\hat{H}}{\hbar},\rho \bigg] + \mathcal{D}_{\rho}[\hat{L}],
\end{equation}
where
\begin{equation}
    \hat{L}=\sqrt{\gamma_{ig}}\ket{g}\bra{i}, \qquad \mathcal{D}_\rho[\hat{O}] \!=\! \hat{O} \rho \hat{O}^\dagger \!-\! \tfrac12 \{ \hat{O}^\dagger \hat{O}, \rho \}. 
\end{equation}
A sketch of the considered three-level scheme is shown in Fig.~\ref{app:sketch}. To lighten the notation, from now on we use $\hat{H}$ to indicate $\hat{H}/\hbar$, thus adopting units of $\hbar=1$.

To recover an effective two-level description, we follow the approach proposed in Ref.~\cite{ReiterEffective2012} and 
define the two-dimensional subspace $\mathcal{G}$ generated by $\ket{g}$ and $\ket{r}$, while $\mathcal{E}$ is the one-dimensional intermediate state $\ket{i}$ subspace. The Hamiltonian~(\ref{eq:H_rot}) can be written as
\begin{equation}
\hat{H}=\hat{H}_{\mathcal{G}}+\hat{H}_{\mathcal{E}}+\hat{V}_{+}+\hat{V}_{-},
\end{equation}
where $\hat{H}_{\mathcal{G}}$ and $\hat{H}_{\mathcal{E}}$ are, respectively, the Hamiltonians of the $\mathcal{G}$ and $\mathcal{E}$ subspaces, while $\hat{V}_{+}=\hat{V}_{-}^{\dagger}$ denotes the coupling between the two subspaces, that is assumed to be perturbative. The dissipation describes the decay from the $\mathcal{E}$ subspace to the $\mathcal{G}$ subspace, thus it can be also written as $\hat{L}=\hat{P}_{\mathcal{G}}\hat{L}\hat{P}_{\mathcal{E}}$, where $\hat{P}_{\mathcal{E}}$ and $\hat{P}_{\mathcal{G}}$ are the projectors on $\mathcal{E}$ and $\mathcal{G}$. 

In general, it is possible to find an effective description of the dynamics in the $\mathcal{G}$ subspace~\cite{ReiterEffective2012}. To do so, we first diagonalize the $\mathcal{G}$ Hamiltonian
\begin{equation}
    \hat{H}_{\mathcal{G}}=\sum_{l=1}^{2} E_l^{\mathcal{G}}\ket{E_l^{\mathcal{G}}}\bra{E_l^{\mathcal{G}}},
\end{equation}
where $\{E_l^{\mathcal{G}}\}$ are its eigenvalues and $\{\ket{E_l^{\mathcal{G}}}\}$ the dressed energies. We also decompose $\hat{V}_{\pm}$ using the $\hat{H}_{\mathcal{G}}$ projectors:
\begin{equation}
\hat{V}_{\pm} \equiv \hat{V}_{\pm}^{(1)} +\hat{V}_{\pm}^{(2)} = \sum_{l=1}^2 \hat V_{\pm} \ket{E_l^{\mathcal{G}}}\bra{E_l^{\mathcal{G}}} .
\end{equation}
Moreover, we introduce the non-Hermitian Hamiltonian 
\begin{equation}
\hat{H}_{\rm NH} = \hat{H}_{\mathcal{E}} - \dfrac{i}{2}\hat{L}^{\dagger}\hat{L},    
\end{equation}
acting on the $\mathcal{E}$ subspace. The dynamics of the system can be described by the effective Hamiltonian and Lindblad operators~\cite{ReiterEffective2012}
\begin{align}
&\hat{H}_{\rm eff} = -\frac{1}{2} \bigg[ \hat{V}_- \sum_{l=1}^{2} \left(\hat{H}_{\rm NH}^{(l)}\right)^{-1} \hat{V}_+^{(l)} + {\rm H.c.} \bigg] + \hat{H}_{\mathcal{G}}, 
\label{eq:H_eff_formula}\\&\hat{L}_{\rm eff} = \hat{L} \sum_{l=1}^{2} \left(\hat{H}_{\rm NH}^{(l)}\right)^{-1} \hat{V}_+^{(l)},
\label{eq:L_eff_formula}
\end{align}
where
\begin{equation}
\left(\hat{H}_{\rm NH}^{(l)}\right)^{-1} = \left(\hat{H}_{\rm NH} - E_l^{\mathcal{G}}\right)^{-1}.  
\end{equation}

We now specialize to the case in which the dissipation field is resonant ($\delta = 0$) and we extract an effective description for $\Delta, \Omega_{\rm exc} > 0$. The eigenvalues and eigenvectors of the $\mathcal{G}$ subspace Hamiltonian $\hat{H}_{\rm \mathcal{G}}=\Delta\ket{r}\bra{r}+\frac{\Omega_{\rm exc}}{2}(\ket{r}\bra{g} + \textrm{H.c.})$ are
\begin{subequations}
\begin{align}
&E_1^{\mathcal{G}} = \tfrac12 (\Delta - \Tilde{\Omega}_{\rm exc}),
\quad
\ket{E_1^{\mathcal{G}}} =c_1^g \ket{g} + c_1^r \ket{r},
\\&E_2^{\mathcal{G}} = \tfrac12 (\Delta + \Tilde{\Omega}_{\rm exc}),
\quad
\ket{E_2^{\mathcal{G}}} =c_2^g \ket{g} + c_2^r \ket{r},
\end{align}
\end{subequations}
where $\Tilde{\Omega}_{\rm exc} \coloneqq \sqrt{\Delta^2 + \Omega_{\rm exc}^2}$ and
\begin{subequations}
\begin{align}
&c_1^g = -\Omega_{\rm exc} \big[ \Omega_{\rm exc}^2 + (\Delta - \Tilde{\Omega}_{\rm exc})^2 \big]^{-1/2},
\\&c_2^g = -(\Delta - \Tilde{\Omega}_{\rm exc})\big[ \Omega_{\rm exc}^2 + (\Delta - \Tilde{\Omega}_{\rm exc})^2 \big]^{-1/2},
\\&
c_1^r = \Omega_{\rm exc} \big[ \Omega_{\rm exc}^2 + (\Delta + \Tilde{\Omega}_{\rm exc})^2 \big]^{-1/2},
\\& c_2^r = (\Delta + \Tilde{\Omega}_{\rm exc})\big[ \Omega_{\rm exc}^2 + (\Delta + \Tilde{\Omega}_{\rm exc})^2 \big]^{-1/2}.
\end{align}
\end{subequations}
Using Eqs.~(\ref{eq:H_eff_formula})-(\ref{eq:L_eff_formula}), we obtain the effective Hamiltonian and Lindblad jump operator that can be used to describe the dynamics inside the $\mathcal{G}$ space:
\begin{align}
\label{eq:H2lev}
&\hat{H}_{\rm eff} = \Delta_{\rm eff} \ket{r}\bra{r} + \tfrac12 \left( \Omega_{\rm exc}^{\rm eff} \ket{r}\bra{g} + {\rm H.c.} \right),
\\&\hat{L}_{\rm eff} = L_{\rm eff}^{gg} \ket{g}\bra{g} + L_{\rm eff}^{gr} \ket{g}\bra{r},
\label{eq:effective_twolevel_L}
\end{align}
where $\Omega_{\rm exc}^{\rm eff}=\Omega_{\rm exc} + \Omega_{\rm exc}^{\rm corr}$ and $\Delta_{\rm eff}=\Delta + \Delta_{\rm corr}$, with
\begin{subequations}
\begin{align}
&\Omega_{\rm exc}^{\rm corr}=- \dfrac{\Omega_{\rm diss}^2}{4} (h^{(1)}c_1^r (c_1^g)^* + h^{(2)}c_2^r (c_2^g)^*),
\\&\Delta_{\rm corr} = - \dfrac{\Omega_{\rm diss}^2}{8} \! \left[ (h^{(1)} + \textrm{H.c.}) |c_1^r|^2 + (h^{(2)} +  \textrm{H.c.}) |c_2^r|^2\right],
\\&
L_{\rm eff}^{gr} = \dfrac{\sqrt{\gamma_{ig}}\Omega_{\rm diss}}{2} \left( h^{(1)} |c_1^r|^2 + h^{(2)} |c_2^r|^2 \right),
\\&
L_{\rm eff}^{gg} = \dfrac{\sqrt{\gamma_{ig}}\Omega_{\rm diss}}{2} \left( h^{(1)} c_1^r (c_1^g)^* + h^{(2)} c_2^r (c_2^g)^*  \right),
\end{align}
\end{subequations}
where $h^{(1)}$ and $h^{(2)}$ are defined through
\begin{subequations}
\begin{align}
&\left(\hat{H}_{\rm NH}^{(1)}\right)^{-1} = \dfrac{2}{\Delta + \Tilde{\Omega}_{\rm exc} - i \gamma_{ig}}\ket{i}\bra{i}\coloneqq h^{(1)}\ket{i}\bra{i}, 
\quad
\\&\left(\hat{H}_{\rm NH}^{(2)}\right)^{-1} = \dfrac{2}{\Delta - \Tilde{\Omega}_{\rm exc} - i \gamma_{ig}}\ket{i}\bra{i}\coloneqq h^{(2)}\ket{i}\bra{i}.
\end{align}
\end{subequations}

The value of the effective parameters is in general complex and strongly depends on the original Hamiltonian parameters. However, in some specific limits, the dynamics reduces to a simpler description.
For instance, in the limit $\Omega_{\rm exc}=0$, we have 
\begin{equation}
L_{\rm eff}^{gr} (\Omega_{\rm exc} \!= \!0) =  i\dfrac{\Omega_{\rm diss}}{\sqrt{\gamma_{ig}}}    
\end{equation}
and $L_{\rm eff}^{gg} (\Omega_{\rm exc} \!= \!0)=\Omega_{\rm exc}^{\rm corr}(\Omega_{\rm exc} \!= \!0)=\Delta_{\rm exc}^{\rm corr}(\Omega_{\rm exc} \!= \!0)=0$. Thus, in the limit of zero excitation Rabi frequency, the dynamics is characterized only by the decay of excitations in the Rydberg state to the ground state, with rate
\begin{equation}
\Gamma_{\rm diss} (\Omega_{\rm exc}\!= \!0) = \big| \braket{g|\hat{L}_{\rm eff} |r} \big|^2 = \dfrac{\Omega_{\rm diss}^2}{\gamma_{ig}}.    
\end{equation}

We can also explore the resonant limit $\Delta=0$. The effective parameters are
\begin{subequations}
\begin{eqnarray}
\Omega_{\rm exc}^{\rm eff} (\Delta\!=\!0) & = & \Omega_{\rm exc} \!+ \!\dfrac{1}{2}\!\left(\dfrac{\Omega_{\rm diss}}{\sqrt{\gamma_{ig}}}\right)^{\!\!2} \!\dfrac{\Omega_{\rm exc}}{\gamma_{ig}}\dfrac{1}{1 + \frac{\Omega_{\rm exc}^2}{\gamma_{ig}^2}}, \qquad \quad
\\\Delta_{\rm eff} (\Delta \!= \!0) & = & 0,
\\L_{\rm eff}^{gr} (\Delta \!= \!0) & = & i\dfrac{\Omega_{\rm diss}}{\sqrt{\gamma_{ig}}} \dfrac{1}{1 + \frac{\Omega_{\rm exc}^2}{\gamma_{ig}^2}},
\\L_{\rm eff}^{gg}(\Delta\!=\!0) & = & -\dfrac{\Omega_{\rm diss}}{\sqrt{\gamma_{ig}}} \dfrac{\Omega_{\rm exc}}{\gamma_{ig}}\dfrac{1}{1 + \frac{\Omega_{\rm exc}^2}{\gamma_{ig}^2}}.
\end{eqnarray}
\end{subequations}
In the limit $\Omega_{\rm exc}\ll \gamma_{ig}$, we have $L_{\rm eff}^{gg}(\Delta\!=\!0)\ll L_{\rm eff}^{gr}(\Delta\!=\!0)$ and $\Omega_{\rm exc}^{\rm eff} (\Delta\!=\!0) \approx \Omega_{\rm exc}$, thus the dynamics can be described by
\begin{eqnarray}
\hat{H}_{\rm eff}(\Delta\!=\!0) & \approx & \dfrac{\Omega_{\rm exc}}{2} \left( \ket{r}\bra{g} + \rm H.c.  \right),\label{eq:H_eff_reso}
\\\hat{L}_{\rm eff} (\Delta \!=\! 0) & \approx & i\dfrac{\Omega_{\rm diss}}{\sqrt{\gamma_{ig}}} \ket{g}\bra{r}.
\end{eqnarray}
For this reason, we can identify an effective decay rate
\begin{equation}
\Gamma_{\rm diss} (\Delta\!= \!0) = \big| \braket{g|\hat{L}_{\rm eff} |r} \big|^2 \approx \dfrac{\Omega_{\rm diss}^2}{\gamma_{ig}}.
\label{eq:gamma_eff_reso}
\end{equation}

Summarizing, we reduced the dynamics of the three-level system to an effective two-level system in which the decay from the Rydberg to the ground state can be controlled by the original Hamiltonian parameters, including the intermediate-Rydberg coupling Rabi frequency, the intermediate-ground decay rate, and the detuning. Thus, the physical important features like the decay rate and the steady state population can be controlled by properly choosing the intermediate-Rydberg coupling laser and the intermediate state used to engineer the dissipation.

\section{Steady state in the resonant limit}
\label{sec:Steady}

Here, we compute the steady state of the three-level system and the effective two-level system in the resonant limit considered in the main text. Given the Hamiltonians~(\ref{eq:H_rot}) with intermediate-ground decay $\gamma_{ig}$ and~(\ref{eq:H_eff_reso}) with decay~(\ref{eq:gamma_eff_reso}) for the three- and two-level systems, it is possible to obtain the equations of motion:
\begin{eqnarray}
   d_t \rho_{gg} & = & i\dfrac{\Omega_{\rm exc}}{2}\left(\rho_{gr}-\rho_{rg} \right) + \gamma_{ig}\rho_{ii}, \nonumber \\
    d_t \rho_{ii} & = &i\dfrac{\Omega_{\rm diss}}{2}\left(\rho_{ir}-\rho_{ri} \right) - \gamma_{ig}\rho_{ii}, \nonumber \\
  d_t \rho_{rr} & = & -i\dfrac{\Omega_{\rm diss}}{2}\left(\rho_{ir}-\rho_{ri} \right) - i\dfrac{\Omega_{\rm exc}}{2}\left(\rho_{gr}-\rho_{rg} \right), \nonumber \\
      d_t \rho_{gi} & = &i\dfrac{\Omega_{\rm diss}}{2}\rho_{gr} -i\dfrac{\Omega_{\rm exc}}{2}\rho_{ri}-\dfrac{\gamma_{ig}}{2}\rho_{gi}, \nonumber \\
    d_t \rho_{ig} & = &-i\dfrac{\Omega_{\rm diss}}{2}\rho_{rg} + i\dfrac{\Omega_{\rm exc}}{2}\rho_{ir}-\dfrac{\gamma_{ig}}{2}\rho_{ig}, \nonumber \\
    d_t \rho_{ir} & = & -i\dfrac{\Omega_{\rm diss}}{2}(\rho_{rr}-\rho_{ii})+i\dfrac{\Omega_{\rm exc}}{2}\rho_{ig}-\dfrac{\gamma_{ig}}{2}\rho_{ir}, \nonumber \\
    d_t \rho_{ri} & = & i\dfrac{\Omega_{\rm diss}}{2}(\rho_{rr}-\rho_{ii})-i\dfrac{\Omega_{\rm exc}}{2}\rho_{gi}-\dfrac{\gamma_{ig}}{2}\rho_{ri}, \nonumber \\
     d_t \rho_{gr} & = & i\dfrac{\Omega_{\rm diss}}{2}\rho_{gi} - i\dfrac{\Omega_{\rm exc}}{2}(\rho_{rr}-\rho_{gg}), \nonumber  \\
    d_t \rho_{rg} & = & -i\dfrac{\Omega_{\rm diss}}{2}\rho_{ig} + i\dfrac{\Omega_{\rm exc}}{2}(\rho_{rr}-\rho_{gg}).
\end{eqnarray}
and
\begin{eqnarray}
    d_t \rho_{gg} & = & i\dfrac{\Omega_{\rm exc}}{2}\left(\rho_{gr}-\rho_{rg} \right) + \Gamma_{\rm diss}\rho_{rr}, \nonumber \\
    d_t \rho_{rr} & = & -i\dfrac{\Omega_{\rm exc}}{2}\left(\rho_{gr}-\rho_{rg} \right) - \Gamma_{\rm diss}\rho_{rr}, \nonumber \\
    d_t \rho_{gr} & = & i\dfrac{\Omega_{\rm exc}}{2}\left(\rho_{gg}-\rho_{rr} \right) - \dfrac{\Gamma_{\rm diss}}{2}\rho_{gr}, \nonumber \\
    d_t \rho_{rg} & = & -i\dfrac{\Omega_{\rm exc}}{2}\left(\rho_{gg}-\rho_{rr} \right) - \dfrac{\Gamma_{\rm diss}}{2} \rho_{rg} , \quad
\end{eqnarray}
where the two sets of equations are, respectively, for the three-level and for the two-level system.

Let us first discuss the effective two-level system. By imposing $d_t\rho_{jk}=0$ for each $j,k$, we obtain the steady-state. In particular, its Rydberg-state population is
\begin{equation}
    \rho_{rr,\rm 2lev}^{SS}=\dfrac{\Omega_{\rm exc}^2}{\Gamma_{\rm eff}^2 + 2\Omega_{\rm exc}^2}=1-\rho_{gg,\rm 2lev}^{SS},
\end{equation}
which, in terms of original Hamiltonian parameters, reads
\begin{equation}
\rho_{rr,\rm 2lev}^{SS}=\dfrac{\Omega_{\rm exc}^2}{(\Omega_{\rm diss}^2/\gamma_{ig})^2 + 2\Omega_{\rm exc}^2}.
\label{app:2resSS}
\end{equation}

\begin{figure}[!t]
    \centering
\includegraphics[width=0.8\columnwidth]{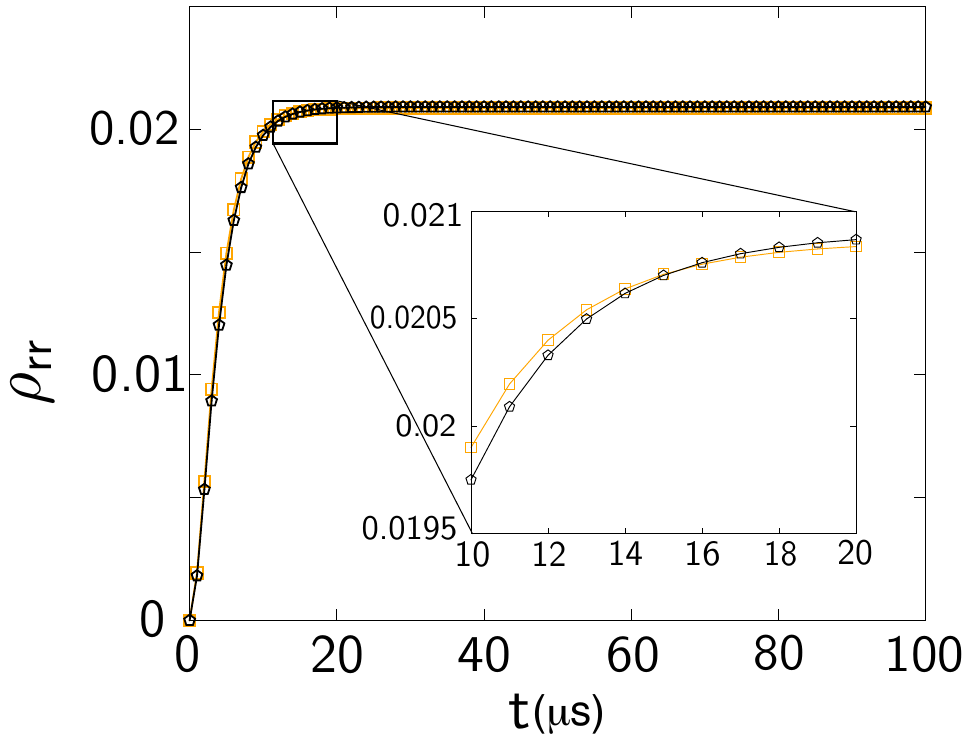}
    \caption{Rydberg-state population as a function of time, for $\Omega_{\rm diss} = 2\pi\times 0.239$ MHz, $\Omega_{\rm exc}=2\pi\times 16$ kHz, $\gamma_{ig}=7$MHz. Both lasers are resonant ($\Delta=\delta=0$) and the pure dephasing rate on the Rydberg-ground transition is $\kappa_{gr}=1$MHz. The plot reports a comparison between simulation results using a three-level (orange squares) and a two-level (black pentagons) description. The inset magnifies the comparison between the two simulations in the time interval $t\in [10,20]$ $\mu$s.}
    \label{app:rr_res}
\end{figure}
\begin{figure*}[!t]
    \centering
\includegraphics[width=\textwidth]{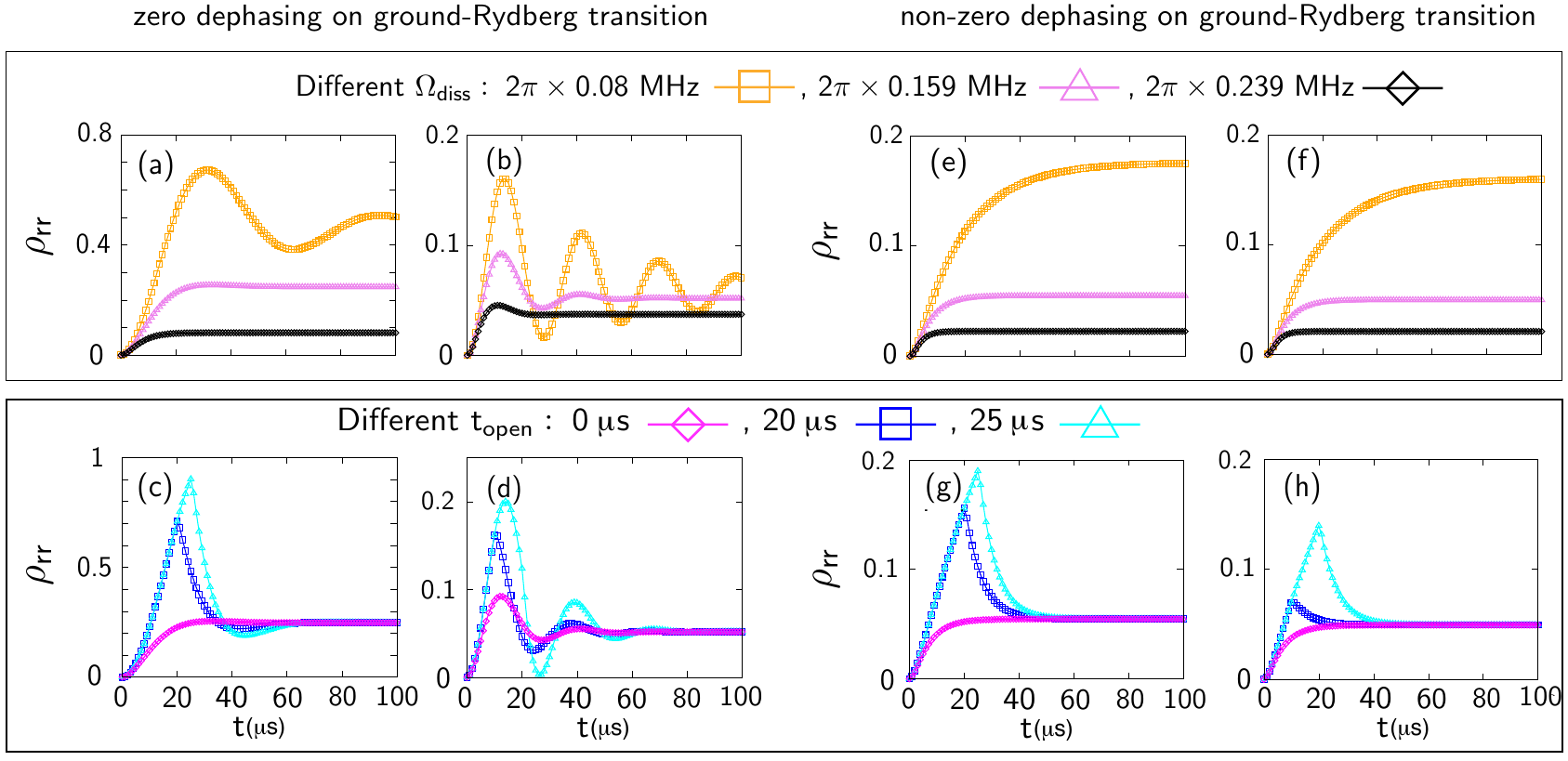}
    \caption{Numerical results for the Rydberg-state population dynamics using the three-level model~(\ref{eq:H_rot}). Panels (a,b,c,d) report the dynamics in the absence of dephasing $\kappa_{gr}=0$, while panels (e,f,g,h) are with $\kappa_{gr}=1$ MHz. In (a,e) the dissipation field $\Omega_{\rm diss}$ is switched on at each time and different values of it are considered ($\Omega_{\rm diss} = 2\pi\times 0.08$, $2\pi\times0.159$, $2\pi\times0.239$ MHz, represented by orange squares, violet triangles, and black rhombuses, respectively), while the lasers are resonant ($\delta=\Delta=0$). (b,f) are the analogous of (a,e), but in the off-resonant case with $\delta=0$ and $\Delta=2\pi\times0.032$ MHz. In (c,g) a dissipation field $\Omega_{\rm diss}=2\pi\times 0.239$MHz is switched on at different times $t_{\rm open}=0$, $20$, $25$ $\mu$s (magenta rhombuses, blue squares, and cyan triangles, respectively), while the lasers are resonant. (d,h) is the analogous of (c,g), with different choices of $t_{\rm open} = 0$, $10$, $20$ $\mu$s (magenta rhombuses, blue squares, and cyan triangles, respectively), but in the off-resonant case $\delta=0$, $\Delta=2\pi\times0.032$ MHz. In all plots we set $\Omega_{\rm exc}=2\pi \times 16$ kHz and $\gamma_{ig}=7$ MHz.}
    \label{app:NumDiss}
\end{figure*}

Coming to the three-level system description, as long as $\gamma_{ig}\gg \Omega_{\rm diss}$, one can see that the intermediate state is weakly populated.
Thus, we first solve the steady-state equations of motion by imposing $d_t \rho_{jk}=0$ and then suppose that the intermediate state population is negligible, meaning that $\rho_{rr}\approx 1-\rho_{gg}$. The resulting steady-state population for $\Omega_{\rm diss}, \Omega_{\rm exc} \neq 0$ is
\begin{equation}
   \rho_{rr,\rm 3lev}^{SS}\approx \dfrac{\Omega_{\rm exc}^2}{\dfrac{\Omega_{\rm diss}^4}{\gamma_{ig}^2}\left(\dfrac{1-\Omega_{\rm exc}^2/\Omega_{\rm diss}^2}{1+\Omega_{\rm exc}^2/\gamma_{ig}^2} \right) + 2\Omega_{\rm exc}^2}.
    \label{app:3resSS}
\end{equation}
In the considered experiment $\Omega_{\rm exc}^2\ll \Omega_{\rm diss}^2, \gamma_{ig}^2$, so that the two expressions in~\eqref{app:2resSS} and~\eqref{app:3resSS} for the steady state roughly coincide.

Finally, we consider the possible presence of pure dephasing in the $\ket{g} \leftrightarrow \ket{r}$ transition. In the two-level system, it can described by adding a new dissipation term $L_{\rm deph}=\sqrt{\kappa_{gr}}\ket{r}\bra{r}$ in the master equation. In the three-level description, it corresponds to adding $-(\kappa_{gr}/2) \rho_{gr}$ and $-(\kappa_{gr}/2) \rho_{rg}$ terms respectively to the equations of motion for $d_t \rho_{gr}$ and $d_t\rho_{rg}$. In the resonant regime ($\delta=\Delta=0$), the two- and three-level steady state Rydberg populations are
\begin{eqnarray}
    \rho_{rr,\rm 2lev}^{SS} 
    & = & \dfrac{\Omega_{\rm exc}^2}{\left(\dfrac{\Omega_{\rm diss}^2}{\gamma_{ig}}\right)^2+\dfrac{\Omega_{\rm diss}^2\kappa_{gr}}{\gamma_{ig}}+ 2\Omega_{\rm exc}^2 }, \\
    \rho_{rr,\rm 3lev}^{SS} 
    & \approx & \Omega_{\rm exc}^2 \Bigg\{ \left(\dfrac{\Omega_{\rm diss}^2}{\gamma_{ig}}\right)^2\left(\dfrac{1-\Omega_{\rm exc}^2 / \Omega_{\rm diss}^2}{1+\Omega_{\rm exc}^2 / \gamma_{ig}^2} \right) \nonumber \\
    && + \dfrac{\Omega_{\rm diss}^2\kappa_{gr}}{\gamma_{ig}\left(1+\Omega_{\rm exc}^2 / \gamma_{ig}^2\right)} + 2\Omega_{\rm exc}^2 \Bigg\}^{-1} .
\end{eqnarray}
The dephasing is responsible for the presence of an additive term at the denominator, thus its role is to decrease the Rydberg-state population. In the presence of dephasing and in the regime considered in the experiment ($\Omega_{\rm exc}^2\ll \Omega_{\rm diss}^2, \gamma_{ig}^2$), the two- and three-level expressions roughly coincide.

\section{Numerical simulations}
\label{sec:numerical}

To witness the validity of the effective two-level description, we resort to numerical simulations performed using \texttt{QuantumOptics.jl}~\cite{Kramer2018quantum}. We report results in the resonant and in the off-resonant cases. We also consider the possibility to have a time-dependent dissipation and pure dephasing on the Rydberg-ground transition. We stress that the off-resonant case is not of interest for the considered experiment, even if the two-level description remains valid also in this regime.

Switching on of the dissipation laser $\Omega_{\rm diss}$ corresponds to switching on the dissipation $\Gamma_{\rm diss}$. To be more precise, we use the following form of the intermediate-Rydberg state coupling:
\begin{equation}
    \Omega_{\rm diss}(t)=\Omega_{\rm diss}
 \: \theta (t-t_{\rm open}),
\end{equation}
where $\theta(t)$ denotes the Heaviside step function. Since $L_{\rm eff}^{gr},L_{\rm eff}^{gg}\propto \Omega_{\rm diss}$ and $\Omega_{\rm exc}^{\rm corr},\Delta_{\rm corr}\propto \Omega_{\rm diss}^2$, the effective dynamics is described by~\eqref{eq:H2lev} and~\eqref{eq:effective_twolevel_L} with time dependent parameters
\begin{subequations}
\begin{align}
& L_{\rm eff}^{gr}(t) = L_{\rm eff}^{gr}  \:\theta (t-t_{\rm open}), 
\\&L_{\rm eff}^{gg}(t) = L_{\rm eff}^{gg}  \:\theta (t-t_{\rm open}),
\\&\Delta_{\rm eff}(t) = \Delta + \Delta_{\rm corr}  \:\theta (t-t_{\rm open}),
\\&\Omega_{\rm exc}^{\rm eff}(t) = \Omega_{\rm exc} + \Omega_{\rm exc}^{\rm corr} \:\theta (t-t_{\rm open}). 
\end{align}
\end{subequations}
Figure~\ref{app:rr_res} reports the Rydberg-state population dynamics for fixed values of the dissipation field Rabi frequency $\Omega_{\rm diss}$, the intermediate-ground decay rate $\gamma_{ig}$, the Rydberg-ground dephasing rate $\kappa_{gr}$ and the excitation field Rabi frequency $\Omega_{\rm exc}$. We clearly observe that the simulations obtained with the two- and the three-level descriptions are well superposed, indicating that the effective model works well. The inset reports the Rydberg population dynamics in a narrow region of $\rho_{rr}$ and highlights the tiny difference between the two simulations.

Figure~\ref{app:NumDiss} reports the Rydberg-state population dynamics for both time-dependent and time-independent protocols, in the presence and in the absence of detuning and dephasing. The panels on the left differ from those on the right by the presence of pure dephasing in the Rydberg-ground transition. While in its absence, the population shows clear oscillations, adding dephasing washes out oscillations and the dynamics acquires a behavior qualitatively similar to what observed experimentally and reported in the main text.  In the presence of pure dephasing, we show that the system reaches the steady state in a relatively short time, with respect to the typical relaxation time of Rydberg states. For the particular choice of the parameters shown in the figure, in the resonant regime, the effective decay rates are $\Gamma_{\rm diss} \approx 35$, $143$, $322$ kHz (respectively for $\Omega_{\rm diss}=2\pi\times 0.08$, $2\pi\times0.159$, $2\pi\times0.239$ MHz), and so they are much bigger than the typical one of Rydberg states, which is of the order of $\sim 10$ kHz. We also observe that the time-dependent protocol is a tool to dynamically control the dissipation.


\begin{thebibliography}{43}%
\makeatletter
\providecommand \@ifxundefined [1]{%
 \@ifx{#1\undefined}
}%
\providecommand \@ifnum [1]{%
 \ifnum #1\expandafter \@firstoftwo
 \else \expandafter \@secondoftwo
 \fi
}%
\providecommand \@ifx [1]{%
 \ifx #1\expandafter \@firstoftwo
 \else \expandafter \@secondoftwo
 \fi
}%
\providecommand \natexlab [1]{#1}%
\providecommand \enquote  [1]{``#1''}%
\providecommand \bibnamefont  [1]{#1}%
\providecommand \bibfnamefont [1]{#1}%
\providecommand \citenamefont [1]{#1}%
\providecommand \href@noop [0]{\@secondoftwo}%
\providecommand \href [0]{\begingroup \@sanitize@url \@href}%
\providecommand \@href[1]{\@@startlink{#1}\@@href}%
\providecommand \@@href[1]{\endgroup#1\@@endlink}%
\providecommand \@sanitize@url [0]{\catcode `\\12\catcode `\$12\catcode `\&12\catcode `\#12\catcode `\^12\catcode `\_12\catcode `\%12\relax}%
\providecommand \@@startlink[1]{}%
\providecommand \@@endlink[0]{}%
\providecommand \url  [0]{\begingroup\@sanitize@url \@url }%
\providecommand \@url [1]{\endgroup\@href {#1}{\urlprefix }}%
\providecommand \urlprefix  [0]{URL }%
\providecommand \Eprint [0]{\href }%
\providecommand \doibase [0]{https://doi.org/}%
\providecommand \selectlanguage [0]{\@gobble}%
\providecommand \bibinfo  [0]{\@secondoftwo}%
\providecommand \bibfield  [0]{\@secondoftwo}%
\providecommand \translation [1]{[#1]}%
\providecommand \BibitemOpen [0]{}%
\providecommand \bibitemStop [0]{}%
\providecommand \bibitemNoStop [0]{.\EOS\space}%
\providecommand \EOS [0]{\spacefactor3000\relax}%
\providecommand \BibitemShut  [1]{\csname bibitem#1\endcsname}%
\let\auto@bib@innerbib\@empty
\bibitem [{\citenamefont {Saffman}(2016)}]{Saffman2016}%
  \BibitemOpen
  \bibfield  {author} {\bibinfo {author} {\bibfnamefont {M.}~\bibnamefont {Saffman}},\ }\bibfield  {title} {\bibinfo {title} {Quantum computing with atomic qubits and {Rydberg} interactions: Progress and challenges},\ }\href {https://doi.org/10.1088/0953-4075/49/20/202001} {\bibfield  {journal} {\bibinfo  {journal} {J. Phys. B: At. Mol. Opt. Phys.}\ }\textbf {\bibinfo {volume} {49}},\ \bibinfo {pages} {202001} (\bibinfo {year} {2016})}\BibitemShut {NoStop}%
\bibitem [{\citenamefont {Browaeys}\ and\ \citenamefont {Lahaye}(2020)}]{Browaeys2020132}%
  \BibitemOpen
  \bibfield  {author} {\bibinfo {author} {\bibfnamefont {A.}~\bibnamefont {Browaeys}}\ and\ \bibinfo {author} {\bibfnamefont {T.}~\bibnamefont {Lahaye}},\ }\bibfield  {title} {\bibinfo {title} {Many-body physics with individually controlled {Rydberg} atoms},\ }\href {https://doi.org/10.1038/s41567-019-0733-z} {\bibfield  {journal} {\bibinfo  {journal} {Nat. Phys.}\ }\textbf {\bibinfo {volume} {16}},\ \bibinfo {pages} {132} (\bibinfo {year} {2020})}\BibitemShut {NoStop}%
\bibitem [{\citenamefont {Morgado}\ and\ \citenamefont {Whitlock}(2021)}]{Morgado2021}%
  \BibitemOpen
  \bibfield  {author} {\bibinfo {author} {\bibfnamefont {M.}~\bibnamefont {Morgado}}\ and\ \bibinfo {author} {\bibfnamefont {S.}~\bibnamefont {Whitlock}},\ }\bibfield  {title} {\bibinfo {title} {Quantum simulation and computing with {Rydberg}-interacting qubits},\ }\href {https://doi.org/10.1116/5.0036562} {\bibfield  {journal} {\bibinfo  {journal} {AVS Quantum Sci.}\ }\textbf {\bibinfo {volume} {3}},\ \bibinfo {pages} {023501} (\bibinfo {year} {2021})}\BibitemShut {NoStop}%
\bibitem [{\citenamefont {Gutiérrez}\ \emph {et~al.}(2017)\citenamefont {Gutiérrez}, \citenamefont {Simonelli}, \citenamefont {Archimi}, \citenamefont {Castellucci}, \citenamefont {Arimondo}, \citenamefont {Ciampini}, \citenamefont {Marcuzzi}, \citenamefont {Lesanovsky},\ and\ \citenamefont {Morsch}}]{Gutierrez2017}%
  \BibitemOpen
  \bibfield  {author} {\bibinfo {author} {\bibfnamefont {R.}~\bibnamefont {Gutiérrez}}, \bibinfo {author} {\bibfnamefont {C.}~\bibnamefont {Simonelli}}, \bibinfo {author} {\bibfnamefont {M.}~\bibnamefont {Archimi}}, \bibinfo {author} {\bibfnamefont {F.}~\bibnamefont {Castellucci}}, \bibinfo {author} {\bibfnamefont {E.}~\bibnamefont {Arimondo}}, \bibinfo {author} {\bibfnamefont {D.}~\bibnamefont {Ciampini}}, \bibinfo {author} {\bibfnamefont {M.}~\bibnamefont {Marcuzzi}}, \bibinfo {author} {\bibfnamefont {I.}~\bibnamefont {Lesanovsky}},\ and\ \bibinfo {author} {\bibfnamefont {O.}~\bibnamefont {Morsch}},\ }\bibfield  {title} {\bibinfo {title} {Experimental signatures of an absorbing-state phase transition in an open driven many-body quantum system},\ }\href {https://doi.org/10.1103/PhysRevA.96.041602} {\bibfield  {journal} {\bibinfo  {journal} {Phys. Rev. A}\ }\textbf {\bibinfo {volume} {96}},\ \bibinfo {pages} {041602} (\bibinfo {year} {2017})}\BibitemShut {NoStop}%
\bibitem [{\citenamefont {Morsch}\ and\ \citenamefont {Lesanovsky}(2018)}]{Morsch2018383}%
  \BibitemOpen
  \bibfield  {author} {\bibinfo {author} {\bibfnamefont {O.}~\bibnamefont {Morsch}}\ and\ \bibinfo {author} {\bibfnamefont {I.}~\bibnamefont {Lesanovsky}},\ }\bibfield  {title} {\bibinfo {title} {Dissipative many-body physics of cold {Rydberg} atoms},\ }\href {https://doi.org/10.1393/ncr/i2018-10149-7} {\bibfield  {journal} {\bibinfo  {journal} {Riv. Nuovo Cimento}\ }\textbf {\bibinfo {volume} {41}},\ \bibinfo {pages} {383} (\bibinfo {year} {2018})}\BibitemShut {NoStop}%
\bibitem [{\citenamefont {Helmrich}\ \emph {et~al.}(2020)\citenamefont {Helmrich}, \citenamefont {Arias}, \citenamefont {Lochead}, \citenamefont {Wintermantel}, \citenamefont {Buchhold}, \citenamefont {Diehl},\ and\ \citenamefont {Whitlock}}]{Helmrich2020481}%
  \BibitemOpen
  \bibfield  {author} {\bibinfo {author} {\bibfnamefont {S.}~\bibnamefont {Helmrich}}, \bibinfo {author} {\bibfnamefont {A.}~\bibnamefont {Arias}}, \bibinfo {author} {\bibfnamefont {G.}~\bibnamefont {Lochead}}, \bibinfo {author} {\bibfnamefont {T.}~\bibnamefont {Wintermantel}}, \bibinfo {author} {\bibfnamefont {M.}~\bibnamefont {Buchhold}}, \bibinfo {author} {\bibfnamefont {S.}~\bibnamefont {Diehl}},\ and\ \bibinfo {author} {\bibfnamefont {S.}~\bibnamefont {Whitlock}},\ }\bibfield  {title} {\bibinfo {title} {Signatures of self-organized criticality in an ultracold atomic gas},\ }\href {https://doi.org/10.1038/s41586-019-1908-6} {\bibfield  {journal} {\bibinfo  {journal} {Nature}\ }\textbf {\bibinfo {volume} {577}},\ \bibinfo {pages} {481} (\bibinfo {year} {2020})}\BibitemShut {NoStop}%
\bibitem [{\citenamefont {Brennecke}\ \emph {et~al.}(2013)\citenamefont {Brennecke}, \citenamefont {Mottl}, \citenamefont {Baumann}, \citenamefont {Landig}, \citenamefont {Donner},\ and\ \citenamefont {Esslinger}}]{Brennecke2013}%
  \BibitemOpen
  \bibfield  {author} {\bibinfo {author} {\bibfnamefont {F.}~\bibnamefont {Brennecke}}, \bibinfo {author} {\bibfnamefont {R.}~\bibnamefont {Mottl}}, \bibinfo {author} {\bibfnamefont {K.}~\bibnamefont {Baumann}}, \bibinfo {author} {\bibfnamefont {R.}~\bibnamefont {Landig}}, \bibinfo {author} {\bibfnamefont {T.}~\bibnamefont {Donner}},\ and\ \bibinfo {author} {\bibfnamefont {T.}~\bibnamefont {Esslinger}},\ }\bibfield  {title} {\bibinfo {title} {Real-time observation of fluctuations at the driven-dissipative {Dicke} phase transition},\ }\href {https://doi.org/10.1073/pnas.1306993110} {\bibfield  {journal} {\bibinfo  {journal} {Proc. Natl. Acad. Sci. USA}\ }\textbf {\bibinfo {volume} {110}},\ \bibinfo {pages} {11763} (\bibinfo {year} {2013})}\BibitemShut {NoStop}%
\bibitem [{\citenamefont {Rodriguez}\ \emph {et~al.}(2017)\citenamefont {Rodriguez}, \citenamefont {Casteels}, \citenamefont {Storme}, \citenamefont {Carlon~Zambon}, \citenamefont {Sagnes}, \citenamefont {Le~Gratiet}, \citenamefont {Galopin}, \citenamefont {Lemaître}, \citenamefont {Amo}, \citenamefont {Ciuti},\ and\ \citenamefont {Bloch}}]{Rodriguez2017}%
  \BibitemOpen
  \bibfield  {author} {\bibinfo {author} {\bibfnamefont {S.~R.~K.}\ \bibnamefont {Rodriguez}}, \bibinfo {author} {\bibfnamefont {W.}~\bibnamefont {Casteels}}, \bibinfo {author} {\bibfnamefont {F.}~\bibnamefont {Storme}}, \bibinfo {author} {\bibfnamefont {N.}~\bibnamefont {Carlon~Zambon}}, \bibinfo {author} {\bibfnamefont {I.}~\bibnamefont {Sagnes}}, \bibinfo {author} {\bibfnamefont {L.}~\bibnamefont {Le~Gratiet}}, \bibinfo {author} {\bibfnamefont {E.}~\bibnamefont {Galopin}}, \bibinfo {author} {\bibfnamefont {A.}~\bibnamefont {Lemaître}}, \bibinfo {author} {\bibfnamefont {A.}~\bibnamefont {Amo}}, \bibinfo {author} {\bibfnamefont {C.}~\bibnamefont {Ciuti}},\ and\ \bibinfo {author} {\bibfnamefont {J.}~\bibnamefont {Bloch}},\ }\bibfield  {title} {\bibinfo {title} {Probing a dissipative phase transition via dynamical optical hysteresis},\ }\href {https://doi.org/10.1103/PhysRevLett.118.247402} {\bibfield  {journal} {\bibinfo  {journal} {Phys. Rev. Lett.}\ }\textbf {\bibinfo {volume} {118}},\ \bibinfo {pages}
  {247402} (\bibinfo {year} {2017})}\BibitemShut {NoStop}%
\bibitem [{\citenamefont {Fink}\ \emph {et~al.}(2017)\citenamefont {Fink}, \citenamefont {Dombi}, \citenamefont {Vukics}, \citenamefont {Wallraff},\ and\ \citenamefont {Domokos}}]{Fink2017}%
  \BibitemOpen
  \bibfield  {author} {\bibinfo {author} {\bibfnamefont {J.~M.}\ \bibnamefont {Fink}}, \bibinfo {author} {\bibfnamefont {A.}~\bibnamefont {Dombi}}, \bibinfo {author} {\bibfnamefont {A.}~\bibnamefont {Vukics}}, \bibinfo {author} {\bibfnamefont {A.}~\bibnamefont {Wallraff}},\ and\ \bibinfo {author} {\bibfnamefont {P.}~\bibnamefont {Domokos}},\ }\bibfield  {title} {\bibinfo {title} {Observation of the photon-blockade breakdown phase transition},\ }\href {https://doi.org/10.1103/PhysRevX.7.011012} {\bibfield  {journal} {\bibinfo  {journal} {Phys. Rev. X}\ }\textbf {\bibinfo {volume} {7}},\ \bibinfo {pages} {011012} (\bibinfo {year} {2017})}\BibitemShut {NoStop}%
\bibitem [{\citenamefont {Fink}\ \emph {et~al.}(2018)\citenamefont {Fink}, \citenamefont {Schade}, \citenamefont {H{\"o}fling}, \citenamefont {Schneider},\ and\ \citenamefont {Imamoglu}}]{Fink2017-2}%
  \BibitemOpen
  \bibfield  {author} {\bibinfo {author} {\bibfnamefont {T.}~\bibnamefont {Fink}}, \bibinfo {author} {\bibfnamefont {A.}~\bibnamefont {Schade}}, \bibinfo {author} {\bibfnamefont {S.}~\bibnamefont {H{\"o}fling}}, \bibinfo {author} {\bibfnamefont {C.}~\bibnamefont {Schneider}},\ and\ \bibinfo {author} {\bibfnamefont {A.}~\bibnamefont {Imamoglu}},\ }\bibfield  {title} {\bibinfo {title} {Signatures of a dissipative phase transition in photon correlation measurements},\ }\href {https://doi.org/10.1038/s41567-017-0020-9} {\bibfield  {journal} {\bibinfo  {journal} {Nat. Phys.}\ }\textbf {\bibinfo {volume} {14}},\ \bibinfo {pages} {365} (\bibinfo {year} {2018})}\BibitemShut {NoStop}%
\bibitem [{\citenamefont {Fitzpatrick}\ \emph {et~al.}(2017)\citenamefont {Fitzpatrick}, \citenamefont {Sundaresan}, \citenamefont {Li}, \citenamefont {Koch},\ and\ \citenamefont {Houck}}]{Fitzpatrick2017}%
  \BibitemOpen
  \bibfield  {author} {\bibinfo {author} {\bibfnamefont {M.}~\bibnamefont {Fitzpatrick}}, \bibinfo {author} {\bibfnamefont {N.~M.}\ \bibnamefont {Sundaresan}}, \bibinfo {author} {\bibfnamefont {A.~C.~Y.}\ \bibnamefont {Li}}, \bibinfo {author} {\bibfnamefont {J.}~\bibnamefont {Koch}},\ and\ \bibinfo {author} {\bibfnamefont {A.~A.}\ \bibnamefont {Houck}},\ }\bibfield  {title} {\bibinfo {title} {Observation of a dissipative phase transition in a one-dimensional circuit {QED} lattice},\ }\href {https://doi.org/10.1103/PhysRevX.7.011016} {\bibfield  {journal} {\bibinfo  {journal} {Phys. Rev. X}\ }\textbf {\bibinfo {volume} {7}},\ \bibinfo {pages} {011016} (\bibinfo {year} {2017})}\BibitemShut {NoStop}%
\bibitem [{\citenamefont {Cai}\ \emph {et~al.}(2021)\citenamefont {Cai}, \citenamefont {Liu}, \citenamefont {Zhao}, \citenamefont {Wu}, \citenamefont {Mei}, \citenamefont {Jiang}, \citenamefont {He}, \citenamefont {Zhang}, \citenamefont {Zhou},\ and\ \citenamefont {Duan}}]{Cai2021}%
  \BibitemOpen
  \bibfield  {author} {\bibinfo {author} {\bibfnamefont {M.-L.}\ \bibnamefont {Cai}}, \bibinfo {author} {\bibfnamefont {Z.-D.}\ \bibnamefont {Liu}}, \bibinfo {author} {\bibfnamefont {W.-D.}\ \bibnamefont {Zhao}}, \bibinfo {author} {\bibfnamefont {Y.-K.}\ \bibnamefont {Wu}}, \bibinfo {author} {\bibfnamefont {Q.-X.}\ \bibnamefont {Mei}}, \bibinfo {author} {\bibfnamefont {Y.}~\bibnamefont {Jiang}}, \bibinfo {author} {\bibfnamefont {L.}~\bibnamefont {He}}, \bibinfo {author} {\bibfnamefont {X.}~\bibnamefont {Zhang}}, \bibinfo {author} {\bibfnamefont {Z.-C.}\ \bibnamefont {Zhou}},\ and\ \bibinfo {author} {\bibfnamefont {L.-M.}\ \bibnamefont {Duan}},\ }\bibfield  {title} {\bibinfo {title} {Observation of a quantum phase transition in the quantum {Rabi} model with a single trapped ion},\ }\href {https://doi.org/10.1038/s41467-021-21425-8} {\bibfield  {journal} {\bibinfo  {journal} {Nat. Commun.}\ }\textbf {\bibinfo {volume} {12}},\ \bibinfo {pages} {1126} (\bibinfo {year} {2021})}\BibitemShut {NoStop}%
\bibitem [{\citenamefont {Li}\ \emph {et~al.}(2018)\citenamefont {Li}, \citenamefont {Chen},\ and\ \citenamefont {Fisher}}]{LiZeno2018}%
  \BibitemOpen
  \bibfield  {author} {\bibinfo {author} {\bibfnamefont {Y.}~\bibnamefont {Li}}, \bibinfo {author} {\bibfnamefont {X.}~\bibnamefont {Chen}},\ and\ \bibinfo {author} {\bibfnamefont {M.~P.~A.}\ \bibnamefont {Fisher}},\ }\bibfield  {title} {\bibinfo {title} {Quantum zeno effect and the many-body entanglement transition},\ }\href {https://doi.org/10.1103/PhysRevB.98.205136} {\bibfield  {journal} {\bibinfo  {journal} {Phys. Rev. B}\ }\textbf {\bibinfo {volume} {98}},\ \bibinfo {pages} {205136} (\bibinfo {year} {2018})}\BibitemShut {NoStop}%
\bibitem [{\citenamefont {Turkeshi}\ \emph {et~al.}(2022)\citenamefont {Turkeshi}, \citenamefont {Dalmonte}, \citenamefont {Fazio},\ and\ \citenamefont {Schir\`o}}]{PhysRevB.105.L241114}%
  \BibitemOpen
  \bibfield  {author} {\bibinfo {author} {\bibfnamefont {X.}~\bibnamefont {Turkeshi}}, \bibinfo {author} {\bibfnamefont {M.}~\bibnamefont {Dalmonte}}, \bibinfo {author} {\bibfnamefont {R.}~\bibnamefont {Fazio}},\ and\ \bibinfo {author} {\bibfnamefont {M.}~\bibnamefont {Schir\`o}},\ }\bibfield  {title} {\bibinfo {title} {Entanglement transitions from stochastic resetting of non-hermitian quasiparticles},\ }\href {https://doi.org/10.1103/PhysRevB.105.L241114} {\bibfield  {journal} {\bibinfo  {journal} {Phys. Rev. B}\ }\textbf {\bibinfo {volume} {105}},\ \bibinfo {pages} {L241114} (\bibinfo {year} {2022})}\BibitemShut {NoStop}%
\bibitem [{\citenamefont {Diehl}\ \emph {et~al.}(2008)\citenamefont {Diehl}, \citenamefont {Micheli}, \citenamefont {Kantian}, \citenamefont {Kraus}, \citenamefont {B{\"u}chler},\ and\ \citenamefont {Zoller}}]{Diehl2008}%
  \BibitemOpen
  \bibfield  {author} {\bibinfo {author} {\bibfnamefont {S.}~\bibnamefont {Diehl}}, \bibinfo {author} {\bibfnamefont {A.}~\bibnamefont {Micheli}}, \bibinfo {author} {\bibfnamefont {A.}~\bibnamefont {Kantian}}, \bibinfo {author} {\bibfnamefont {B.}~\bibnamefont {Kraus}}, \bibinfo {author} {\bibfnamefont {H.~P.}\ \bibnamefont {B{\"u}chler}},\ and\ \bibinfo {author} {\bibfnamefont {P.}~\bibnamefont {Zoller}},\ }\bibfield  {title} {\bibinfo {title} {Quantum states and phases in driven open quantum systems with cold atoms},\ }\href {https://doi.org/10.1038/nphys1073} {\bibfield  {journal} {\bibinfo  {journal} {Nat. Phys.}\ }\textbf {\bibinfo {volume} {4}},\ \bibinfo {pages} {878} (\bibinfo {year} {2008})}\BibitemShut {NoStop}%
\bibitem [{\citenamefont {Verstraete}\ \emph {et~al.}(2009)\citenamefont {Verstraete}, \citenamefont {Wolf},\ and\ \citenamefont {Cirac}}]{Verstraete2009}%
  \BibitemOpen
  \bibfield  {author} {\bibinfo {author} {\bibfnamefont {F.}~\bibnamefont {Verstraete}}, \bibinfo {author} {\bibfnamefont {M.~M.}\ \bibnamefont {Wolf}},\ and\ \bibinfo {author} {\bibfnamefont {J.~I.}\ \bibnamefont {Cirac}},\ }\bibfield  {title} {\bibinfo {title} {Quantum computation and quantum-state engineering driven by dissipation},\ }\href {https://doi.org/10.1038/nphys1342} {\bibfield  {journal} {\bibinfo  {journal} {Nat. Phys.}\ }\textbf {\bibinfo {volume} {5}},\ \bibinfo {pages} {633} (\bibinfo {year} {2009})}\BibitemShut {NoStop}%
\bibitem [{\citenamefont {Müller}\ \emph {et~al.}(2012)\citenamefont {Müller}, \citenamefont {Diehl}, \citenamefont {Pupillo},\ and\ \citenamefont {Zoller}}]{Muller20121}%
  \BibitemOpen
  \bibfield  {author} {\bibinfo {author} {\bibfnamefont {M.}~\bibnamefont {Müller}}, \bibinfo {author} {\bibfnamefont {S.}~\bibnamefont {Diehl}}, \bibinfo {author} {\bibfnamefont {G.}~\bibnamefont {Pupillo}},\ and\ \bibinfo {author} {\bibfnamefont {P.}~\bibnamefont {Zoller}},\ }\bibfield  {title} {\bibinfo {title} {Engineered open systems and quantum simulations with atoms and ions},\ }\href {https://doi.org/10.1016/B978-0-12-396482-3.00001-6} {\bibfield  {journal} {\bibinfo  {journal} {Adv. At. Mol. Opt. Phys.}\ }\textbf {\bibinfo {volume} {61}},\ \bibinfo {pages} {1} (\bibinfo {year} {2012})}\BibitemShut {NoStop}%
\bibitem [{\citenamefont {Rao}\ and\ \citenamefont {Mølmer}(2013)}]{Rao2013}%
  \BibitemOpen
  \bibfield  {author} {\bibinfo {author} {\bibfnamefont {D.~D.~B.}\ \bibnamefont {Rao}}\ and\ \bibinfo {author} {\bibfnamefont {K.}~\bibnamefont {Mølmer}},\ }\bibfield  {title} {\bibinfo {title} {Dark entangled steady states of interacting rydberg atoms},\ }\href {https://doi.org/10.1103/PhysRevLett.111.033606} {\bibfield  {journal} {\bibinfo  {journal} {Phys. Rev. Lett.}\ }\textbf {\bibinfo {volume} {111}},\ \bibinfo {pages} {033606} (\bibinfo {year} {2013})}\BibitemShut {NoStop}%
\bibitem [{\citenamefont {Carr}\ and\ \citenamefont {Saffman}(2013)}]{Carr2013}%
  \BibitemOpen
  \bibfield  {author} {\bibinfo {author} {\bibfnamefont {A.}~\bibnamefont {Carr}}\ and\ \bibinfo {author} {\bibfnamefont {M.}~\bibnamefont {Saffman}},\ }\bibfield  {title} {\bibinfo {title} {Preparation of entangled and antiferromagnetic states by dissipative {Rydberg} pumping},\ }\href {https://doi.org/10.1103/PhysRevLett.111.033607} {\bibfield  {journal} {\bibinfo  {journal} {Phys. Rev. Lett.}\ }\textbf {\bibinfo {volume} {111}},\ \bibinfo {pages} {033607} (\bibinfo {year} {2013})}\BibitemShut {NoStop}%
\bibitem [{\citenamefont {Rao}\ and\ \citenamefont {Mølmer}(2014)}]{Rao2014}%
  \BibitemOpen
  \bibfield  {author} {\bibinfo {author} {\bibfnamefont {D.~D.~B.}\ \bibnamefont {Rao}}\ and\ \bibinfo {author} {\bibfnamefont {K.}~\bibnamefont {Mølmer}},\ }\bibfield  {title} {\bibinfo {title} {Deterministic entanglement of {Rydberg} ensembles by engineered dissipation},\ }\href {https://doi.org/10.1103/PhysRevA.90.062319} {\bibfield  {journal} {\bibinfo  {journal} {Phys. Rev. A}\ }\textbf {\bibinfo {volume} {90}},\ \bibinfo {pages} {062319} (\bibinfo {year} {2014})}\BibitemShut {NoStop}%
\bibitem [{\citenamefont {Lee}\ \emph {et~al.}(2015)\citenamefont {Lee}, \citenamefont {Cho},\ and\ \citenamefont {Choi}}]{Lee2015}%
  \BibitemOpen
  \bibfield  {author} {\bibinfo {author} {\bibfnamefont {S.~K.}\ \bibnamefont {Lee}}, \bibinfo {author} {\bibfnamefont {J.}~\bibnamefont {Cho}},\ and\ \bibinfo {author} {\bibfnamefont {K.}~\bibnamefont {Choi}},\ }\bibfield  {title} {\bibinfo {title} {Emergence of stationary many-body entanglement in driven-dissipative {Rydberg} lattice gases},\ }\href {https://doi.org/10.1088/1367-2630/17/11/113053} {\bibfield  {journal} {\bibinfo  {journal} {New J. Phys.}\ }\textbf {\bibinfo {volume} {17}},\ \bibinfo {pages} {113053} (\bibinfo {year} {2015})}\BibitemShut {NoStop}%
\bibitem [{\citenamefont {Su}\ \emph {et~al.}(2015)\citenamefont {Su}, \citenamefont {Guo}, \citenamefont {Wang},\ and\ \citenamefont {Zhang}}]{Su2015}%
  \BibitemOpen
  \bibfield  {author} {\bibinfo {author} {\bibfnamefont {S.-L.}\ \bibnamefont {Su}}, \bibinfo {author} {\bibfnamefont {Q.}~\bibnamefont {Guo}}, \bibinfo {author} {\bibfnamefont {H.-F.}\ \bibnamefont {Wang}},\ and\ \bibinfo {author} {\bibfnamefont {S.}~\bibnamefont {Zhang}},\ }\bibfield  {title} {\bibinfo {title} {Simplified scheme for entanglement preparation with {Rydberg} pumping via dissipation},\ }\href {https://doi.org/10.1103/PhysRevA.92.022328} {\bibfield  {journal} {\bibinfo  {journal} {Phys. Rev. A}\ }\textbf {\bibinfo {volume} {92}},\ \bibinfo {pages} {022328} (\bibinfo {year} {2015})}\BibitemShut {NoStop}%
\bibitem [{\citenamefont {Shao}\ \emph {et~al.}(2017)\citenamefont {Shao}, \citenamefont {Wu}, \citenamefont {Yi},\ and\ \citenamefont {Long}}]{Shao2017}%
  \BibitemOpen
  \bibfield  {author} {\bibinfo {author} {\bibfnamefont {X.}~\bibnamefont {Shao}}, \bibinfo {author} {\bibfnamefont {J.}~\bibnamefont {Wu}}, \bibinfo {author} {\bibfnamefont {X.}~\bibnamefont {Yi}},\ and\ \bibinfo {author} {\bibfnamefont {G.-L.}\ \bibnamefont {Long}},\ }\bibfield  {title} {\bibinfo {title} {Dissipative preparation of steady {Greenberger-Horne-Zeilinger} states for {Rydberg} atoms with quantum {Zeno} dynamics},\ }\href {https://doi.org/10.1103/PhysRevA.96.062315} {\bibfield  {journal} {\bibinfo  {journal} {Phys. Rev. A}\ }\textbf {\bibinfo {volume} {96}},\ \bibinfo {pages} {062315} (\bibinfo {year} {2017})}\BibitemShut {NoStop}%
\bibitem [{\citenamefont {Roghani}\ and\ \citenamefont {Weimer}(2018)}]{Roghani2018}%
  \BibitemOpen
  \bibfield  {author} {\bibinfo {author} {\bibfnamefont {M.}~\bibnamefont {Roghani}}\ and\ \bibinfo {author} {\bibfnamefont {H.}~\bibnamefont {Weimer}},\ }\bibfield  {title} {\bibinfo {title} {Dissipative preparation of entangled many-body states with {Rydberg} atoms},\ }\href {https://doi.org/10.1088/2058-9565/aab3f3} {\bibfield  {journal} {\bibinfo  {journal} {Quantum Sci. Technol.}\ }\textbf {\bibinfo {volume} {3}},\ \bibinfo {pages} {035002} (\bibinfo {year} {2018})}\BibitemShut {NoStop}%
\bibitem [{\citenamefont {Harrington}\ \emph {et~al.}(2022)\citenamefont {Harrington}, \citenamefont {Mueller},\ and\ \citenamefont {Murch}}]{Harrington2022660}%
  \BibitemOpen
  \bibfield  {author} {\bibinfo {author} {\bibfnamefont {P.~M.}\ \bibnamefont {Harrington}}, \bibinfo {author} {\bibfnamefont {E.~J.}\ \bibnamefont {Mueller}},\ and\ \bibinfo {author} {\bibfnamefont {K.~W.}\ \bibnamefont {Murch}},\ }\bibfield  {title} {\bibinfo {title} {Engineered dissipation for quantum information science},\ }\href {https://doi.org/10.1038/s42254-022-00494-8} {\bibfield  {journal} {\bibinfo  {journal} {Nat. Rev. Phys.}\ }\textbf {\bibinfo {volume} {4}},\ \bibinfo {pages} {660} (\bibinfo {year} {2022})}\BibitemShut {NoStop}%
\bibitem [{\citenamefont {Yang}\ \emph {et~al.}(2021)\citenamefont {Yang}, \citenamefont {Li},\ and\ \citenamefont {Shao}}]{Yang2021}%
  \BibitemOpen
  \bibfield  {author} {\bibinfo {author} {\bibfnamefont {C.}~\bibnamefont {Yang}}, \bibinfo {author} {\bibfnamefont {D.-X.}\ \bibnamefont {Li}},\ and\ \bibinfo {author} {\bibfnamefont {X.-Q.}\ \bibnamefont {Shao}},\ }\bibfield  {title} {\bibinfo {title} {Dissipative preparation of multipartite {Greenberger-Horne-Zeilinger} states of {Rydberg} atoms},\ }\href {https://doi.org/10.1088/1674-1056/abd755} {\bibfield  {journal} {\bibinfo  {journal} {Chin. Phys. B}\ }\textbf {\bibinfo {volume} {30}},\ \bibinfo {pages} {023201} (\bibinfo {year} {2021})}\BibitemShut {NoStop}%
\bibitem [{\citenamefont {Beterov}\ \emph {et~al.}(2009)\citenamefont {Beterov}, \citenamefont {Ryabtsev}, \citenamefont {Tretyakov},\ and\ \citenamefont {Entin}}]{Beterov2009}%
  \BibitemOpen
  \bibfield  {author} {\bibinfo {author} {\bibfnamefont {I.}~\bibnamefont {Beterov}}, \bibinfo {author} {\bibfnamefont {I.}~\bibnamefont {Ryabtsev}}, \bibinfo {author} {\bibfnamefont {D.}~\bibnamefont {Tretyakov}},\ and\ \bibinfo {author} {\bibfnamefont {V.}~\bibnamefont {Entin}},\ }\bibfield  {title} {\bibinfo {title} {Quasiclassical calculations of blackbody-radiation-induced depopulation rates and effective lifetimes of {Rydberg nS, nP, and nD} alkali-metal atoms with $n\ensuremath{\le}80$},\ }\href {https://doi.org/10.1103/PhysRevA.79.052504} {\bibfield  {journal} {\bibinfo  {journal} {Phys. Rev. A}\ }\textbf {\bibinfo {volume} {79}},\ \bibinfo {pages} {052504} (\bibinfo {year} {2009})}\BibitemShut {NoStop}%
\bibitem [{\citenamefont {Archimi}\ \emph {et~al.}(2022)\citenamefont {Archimi}, \citenamefont {Ceccanti}, \citenamefont {Distefano}, \citenamefont {Di~Virgilio}, \citenamefont {Franco}, \citenamefont {Greco}, \citenamefont {Simonelli}, \citenamefont {Arimondo}, \citenamefont {Ciampini},\ and\ \citenamefont {Morsch}}]{Archimi2022}%
  \BibitemOpen
  \bibfield  {author} {\bibinfo {author} {\bibfnamefont {M.}~\bibnamefont {Archimi}}, \bibinfo {author} {\bibfnamefont {M.}~\bibnamefont {Ceccanti}}, \bibinfo {author} {\bibfnamefont {M.}~\bibnamefont {Distefano}}, \bibinfo {author} {\bibfnamefont {L.}~\bibnamefont {Di~Virgilio}}, \bibinfo {author} {\bibfnamefont {R.}~\bibnamefont {Franco}}, \bibinfo {author} {\bibfnamefont {A.}~\bibnamefont {Greco}}, \bibinfo {author} {\bibfnamefont {C.}~\bibnamefont {Simonelli}}, \bibinfo {author} {\bibfnamefont {E.}~\bibnamefont {Arimondo}}, \bibinfo {author} {\bibfnamefont {D.}~\bibnamefont {Ciampini}},\ and\ \bibinfo {author} {\bibfnamefont {O.}~\bibnamefont {Morsch}},\ }\bibfield  {title} {\bibinfo {title} {Measurements of blackbody-radiation-induced transition rates between high-lying {S, P, and D Rydberg} levels},\ }\href {https://doi.org/10.1103/PhysRevA.105.063104} {\bibfield  {journal} {\bibinfo  {journal} {Phys. Rev. A}\ }\textbf {\bibinfo {volume} {105}},\ \bibinfo {pages} {063104} (\bibinfo {year}
  {2022})}\BibitemShut {NoStop}%
\bibitem [{\citenamefont {Simonelli}\ \emph {et~al.}(2017)\citenamefont {Simonelli}, \citenamefont {Archimi}, \citenamefont {Asteria}, \citenamefont {Capecchi}, \citenamefont {Masella}, \citenamefont {Arimondo}, \citenamefont {Ciampini},\ and\ \citenamefont {Morsch}}]{Simonelli2017}%
  \BibitemOpen
  \bibfield  {author} {\bibinfo {author} {\bibfnamefont {C.}~\bibnamefont {Simonelli}}, \bibinfo {author} {\bibfnamefont {M.}~\bibnamefont {Archimi}}, \bibinfo {author} {\bibfnamefont {L.}~\bibnamefont {Asteria}}, \bibinfo {author} {\bibfnamefont {D.}~\bibnamefont {Capecchi}}, \bibinfo {author} {\bibfnamefont {G.}~\bibnamefont {Masella}}, \bibinfo {author} {\bibfnamefont {E.}~\bibnamefont {Arimondo}}, \bibinfo {author} {\bibfnamefont {D.}~\bibnamefont {Ciampini}},\ and\ \bibinfo {author} {\bibfnamefont {O.}~\bibnamefont {Morsch}},\ }\bibfield  {title} {\bibinfo {title} {Deexcitation spectroscopy of strongly interacting {Rydberg} gases},\ }\href {https://doi.org/10.1103/PhysRevA.96.043411} {\bibfield  {journal} {\bibinfo  {journal} {Phys. Rev. A}\ }\textbf {\bibinfo {volume} {96}},\ \bibinfo {pages} {043411} (\bibinfo {year} {2017})}\BibitemShut {NoStop}%
\bibitem [{\citenamefont {Comparat}\ and\ \citenamefont {Pillet}(2010)}]{Comparat2010A208}%
  \BibitemOpen
  \bibfield  {author} {\bibinfo {author} {\bibfnamefont {D.}~\bibnamefont {Comparat}}\ and\ \bibinfo {author} {\bibfnamefont {P.}~\bibnamefont {Pillet}},\ }\bibfield  {title} {\bibinfo {title} {Dipole blockade in a cold {Rydberg} atomic sample},\ }\href {https://doi.org/10.1364/JOSAB.27.00A208} {\bibfield  {journal} {\bibinfo  {journal} {J. Opt. Soc. Am. B}\ }\textbf {\bibinfo {volume} {27}},\ \bibinfo {pages} {A208} (\bibinfo {year} {2010})}\BibitemShut {NoStop}%
\bibitem [{\citenamefont {Valado}\ \emph {et~al.}(2016)\citenamefont {Valado}, \citenamefont {Simonelli}, \citenamefont {Hoogerland}, \citenamefont {Lesanovsky}, \citenamefont {Garrahan}, \citenamefont {Arimondo}, \citenamefont {Ciampini},\ and\ \citenamefont {Morsch}}]{Valado2016}%
  \BibitemOpen
  \bibfield  {author} {\bibinfo {author} {\bibfnamefont {M.}~\bibnamefont {Valado}}, \bibinfo {author} {\bibfnamefont {C.}~\bibnamefont {Simonelli}}, \bibinfo {author} {\bibfnamefont {M.}~\bibnamefont {Hoogerland}}, \bibinfo {author} {\bibfnamefont {I.}~\bibnamefont {Lesanovsky}}, \bibinfo {author} {\bibfnamefont {J.}~\bibnamefont {Garrahan}}, \bibinfo {author} {\bibfnamefont {E.}~\bibnamefont {Arimondo}}, \bibinfo {author} {\bibfnamefont {D.}~\bibnamefont {Ciampini}},\ and\ \bibinfo {author} {\bibfnamefont {O.}~\bibnamefont {Morsch}},\ }\bibfield  {title} {\bibinfo {title} {Experimental observation of controllable kinetic constraints in a cold atomic gas},\ }\href {https://doi.org/10.1103/PhysRevA.93.040701} {\bibfield  {journal} {\bibinfo  {journal} {Phys. Rev. A}\ }\textbf {\bibinfo {volume} {93}},\ \bibinfo {pages} {040701} (\bibinfo {year} {2016})}\BibitemShut {NoStop}%
\bibitem [{\citenamefont {Šibalić}\ \emph {et~al.}(2017)\citenamefont {Šibalić}, \citenamefont {Pritchard}, \citenamefont {Adams},\ and\ \citenamefont {Weatherill}}]{Sib2017}%
  \BibitemOpen
  \bibfield  {author} {\bibinfo {author} {\bibfnamefont {N.}~\bibnamefont {Šibalić}}, \bibinfo {author} {\bibfnamefont {J.~D.}\ \bibnamefont {Pritchard}}, \bibinfo {author} {\bibfnamefont {C.~S.}\ \bibnamefont {Adams}},\ and\ \bibinfo {author} {\bibfnamefont {K.~J.}\ \bibnamefont {Weatherill}},\ }\bibfield  {title} {\bibinfo {title} {{ARC}: An open-source library for calculating properties of alkali {Rydberg} atoms},\ }\href {https://doi.org/10.1016/j.cpc.2017.06.015} {\bibfield  {journal} {\bibinfo  {journal} {Comput. Phys. Commun.}\ }\textbf {\bibinfo {volume} {220}},\ \bibinfo {pages} {319} (\bibinfo {year} {2017})}\BibitemShut {NoStop}%
\bibitem [{\citenamefont {Nalbach}\ \emph {et~al.}(2015)\citenamefont {Nalbach}, \citenamefont {Vishveshwara},\ and\ \citenamefont {Clerk}}]{Nalbach2015}%
  \BibitemOpen
  \bibfield  {author} {\bibinfo {author} {\bibfnamefont {P.}~\bibnamefont {Nalbach}}, \bibinfo {author} {\bibfnamefont {S.}~\bibnamefont {Vishveshwara}},\ and\ \bibinfo {author} {\bibfnamefont {A.~A.}\ \bibnamefont {Clerk}},\ }\bibfield  {title} {\bibinfo {title} {Quantum {Kibble-Zurek} physics in the presence of spatially correlated dissipation},\ }\href {https://doi.org/10.1103/PhysRevB.92.014306} {\bibfield  {journal} {\bibinfo  {journal} {Phys. Rev. B}\ }\textbf {\bibinfo {volume} {92}},\ \bibinfo {pages} {014306} (\bibinfo {year} {2015})}\BibitemShut {NoStop}%
\bibitem [{\citenamefont {Duan}\ and\ \citenamefont {Guo}(1997)}]{Duan19974466}%
  \BibitemOpen
  \bibfield  {author} {\bibinfo {author} {\bibfnamefont {L.-M.}\ \bibnamefont {Duan}}\ and\ \bibinfo {author} {\bibfnamefont {G.-C.}\ \bibnamefont {Guo}},\ }\bibfield  {title} {\bibinfo {title} {Perturbative expansions for the fidelities and spatially correlated dissipation of quantum bits},\ }\href {https://doi.org/10.1103/PhysRevA.56.4466} {\bibfield  {journal} {\bibinfo  {journal} {Phys. Rev. A}\ }\textbf {\bibinfo {volume} {56}},\ \bibinfo {pages} {4466} (\bibinfo {year} {1997})}\BibitemShut {NoStop}%
\bibitem [{\citenamefont {Masson}\ and\ \citenamefont {Asenjo-Garcia}(2022)}]{Masson2022}%
  \BibitemOpen
  \bibfield  {author} {\bibinfo {author} {\bibfnamefont {S.~J.}\ \bibnamefont {Masson}}\ and\ \bibinfo {author} {\bibfnamefont {A.}~\bibnamefont {Asenjo-Garcia}},\ }\bibfield  {title} {\bibinfo {title} {Universality of {Dicke} superradiance in arrays of quantum emitters},\ }\href {https://doi.org/10.1038/s41467-022-29805-4} {\bibfield  {journal} {\bibinfo  {journal} {Nat. Commun.}\ }\textbf {\bibinfo {volume} {13}},\ \bibinfo {pages} {2285} (\bibinfo {year} {2022})}\BibitemShut {NoStop}%
\bibitem [{\citenamefont {Chakrabarti}\ and\ \citenamefont {Bhattacharyya}(2023)}]{Chakrabarti2023}%
  \BibitemOpen
  \bibfield  {author} {\bibinfo {author} {\bibfnamefont {A.}~\bibnamefont {Chakrabarti}}\ and\ \bibinfo {author} {\bibfnamefont {R.}~\bibnamefont {Bhattacharyya}},\ }\bibfield  {title} {\bibinfo {title} {Emergence of prethermal states in a driven dissipative system through cross-correlated dissipation},\ }\href {https://doi.org/10.1209/0295-5075/acd4e5} {\bibfield  {journal} {\bibinfo  {journal} {Europhys. Lett.}\ }\textbf {\bibinfo {volume} {142}},\ \bibinfo {pages} {55001} (\bibinfo {year} {2023})}\BibitemShut {NoStop}%
\bibitem [{\citenamefont {Kazemi}\ and\ \citenamefont {Weimer}(2023)}]{kazemi2023driven}%
  \BibitemOpen
  \bibfield  {author} {\bibinfo {author} {\bibfnamefont {J.}~\bibnamefont {Kazemi}}\ and\ \bibinfo {author} {\bibfnamefont {H.}~\bibnamefont {Weimer}},\ }\bibfield  {title} {\bibinfo {title} {Driven-dissipative {Rydberg} blockade in optical lattices},\ }\href {https://doi.org/10.1103/PhysRevLett.130.163601} {\bibfield  {journal} {\bibinfo  {journal} {Phys. Rev. Lett.}\ }\textbf {\bibinfo {volume} {130}},\ \bibinfo {pages} {163601} (\bibinfo {year} {2023})}\BibitemShut {NoStop}%
\bibitem [{\citenamefont {Kitson}\ \emph {et~al.}(2024)\citenamefont {Kitson}, \citenamefont {Haug}, \citenamefont {La~Magna}, \citenamefont {Morsch},\ and\ \citenamefont {Amico}}]{kitson2023rydberg}%
  \BibitemOpen
  \bibfield  {author} {\bibinfo {author} {\bibfnamefont {P.}~\bibnamefont {Kitson}}, \bibinfo {author} {\bibfnamefont {T.}~\bibnamefont {Haug}}, \bibinfo {author} {\bibfnamefont {A.}~\bibnamefont {La~Magna}}, \bibinfo {author} {\bibfnamefont {O.}~\bibnamefont {Morsch}},\ and\ \bibinfo {author} {\bibfnamefont {L.}~\bibnamefont {Amico}},\ }\bibfield  {title} {\bibinfo {title} {Rydberg atomtronic devices},\ }\href {https://doi.org/10.1103/PhysRevA.110.043304} {\bibfield  {journal} {\bibinfo  {journal} {Phys. Rev. A}\ }\textbf {\bibinfo {volume} {110}},\ \bibinfo {pages} {043304} (\bibinfo {year} {2024})}\BibitemShut {NoStop}%
\bibitem [{\citenamefont {Reiter}\ and\ \citenamefont {S\o{}rensen}(2012)}]{ReiterEffective2012}%
  \BibitemOpen
  \bibfield  {author} {\bibinfo {author} {\bibfnamefont {F.}~\bibnamefont {Reiter}}\ and\ \bibinfo {author} {\bibfnamefont {A.~S.}\ \bibnamefont {S\o{}rensen}},\ }\bibfield  {title} {\bibinfo {title} {Effective operator formalism for open quantum systems},\ }\href {https://doi.org/10.1103/PhysRevA.85.032111} {\bibfield  {journal} {\bibinfo  {journal} {Phys. Rev. A}\ }\textbf {\bibinfo {volume} {85}},\ \bibinfo {pages} {032111} (\bibinfo {year} {2012})}\BibitemShut {NoStop}%
\bibitem [{\citenamefont {Shao}\ \emph {et~al.}(2023)\citenamefont {Shao}, \citenamefont {Liu}, \citenamefont {Xue}, \citenamefont {Mu},\ and\ \citenamefont {Li}}]{XaoHigh2023}%
  \BibitemOpen
  \bibfield  {author} {\bibinfo {author} {\bibfnamefont {X.}~\bibnamefont {Shao}}, \bibinfo {author} {\bibfnamefont {F.}~\bibnamefont {Liu}}, \bibinfo {author} {\bibfnamefont {X.}~\bibnamefont {Xue}}, \bibinfo {author} {\bibfnamefont {W.}~\bibnamefont {Mu}},\ and\ \bibinfo {author} {\bibfnamefont {W.}~\bibnamefont {Li}},\ }\bibfield  {title} {\bibinfo {title} {High-fidelity interconversion between {Greenberger-Horne-Zeilinger} and ${W}$ states through {Floquet-Lindblad} engineering in {Rydberg} atom arrays},\ }\href {https://doi.org/10.1103/PhysRevApplied.20.014014} {\bibfield  {journal} {\bibinfo  {journal} {Phys. Rev. Appl.}\ }\textbf {\bibinfo {volume} {20}},\ \bibinfo {pages} {014014} (\bibinfo {year} {2023})}\BibitemShut {NoStop}%
\bibitem [{\citenamefont {Schempp}\ \emph {et~al.}(2015)\citenamefont {Schempp}, \citenamefont {G\"unter}, \citenamefont {W\"uster}, \citenamefont {Weidem\"uller},\ and\ \citenamefont {Whitlock}}]{Schempp2015transport}%
  \BibitemOpen
  \bibfield  {author} {\bibinfo {author} {\bibfnamefont {H.}~\bibnamefont {Schempp}}, \bibinfo {author} {\bibfnamefont {G.}~\bibnamefont {G\"unter}}, \bibinfo {author} {\bibfnamefont {S.}~\bibnamefont {W\"uster}}, \bibinfo {author} {\bibfnamefont {M.}~\bibnamefont {Weidem\"uller}},\ and\ \bibinfo {author} {\bibfnamefont {S.}~\bibnamefont {Whitlock}},\ }\bibfield  {title} {\bibinfo {title} {Correlated exciton transport in {Rydberg}-dressed-atom spin chains},\ }\href {https://doi.org/10.1103/PhysRevLett.115.093002} {\bibfield  {journal} {\bibinfo  {journal} {Phys. Rev. Lett.}\ }\textbf {\bibinfo {volume} {115}},\ \bibinfo {pages} {093002} (\bibinfo {year} {2015})}\BibitemShut {NoStop}%
\bibitem [{\citenamefont {Breuer}\ and\ \citenamefont {Petruccione}(2007)}]{breuer2002theory}%
  \BibitemOpen
  \bibfield  {author} {\bibinfo {author} {\bibfnamefont {H.-P.}\ \bibnamefont {Breuer}}\ and\ \bibinfo {author} {\bibfnamefont {F.}~\bibnamefont {Petruccione}},\ }\href {https://doi.org/10.1093/acprof:oso/9780199213900.001.0001} {\emph {\bibinfo {title} {The theory of open quantum systems}}}\ (\bibinfo  {publisher} {Oxford University Press, Oxford},\ \bibinfo {year} {2007})\BibitemShut {NoStop}%
\bibitem [{\citenamefont {Kr\"{a}mer}\ \emph {et~al.}(2018)\citenamefont {Kr\"{a}mer}, \citenamefont {Plankensteiner}, \citenamefont {Ostermann},\ and\ \citenamefont {Ritsch}}]{Kramer2018quantum}%
  \BibitemOpen
  \bibfield  {author} {\bibinfo {author} {\bibfnamefont {S.}~\bibnamefont {Kr\"{a}mer}}, \bibinfo {author} {\bibfnamefont {D.}~\bibnamefont {Plankensteiner}}, \bibinfo {author} {\bibfnamefont {L.}~\bibnamefont {Ostermann}},\ and\ \bibinfo {author} {\bibfnamefont {H.}~\bibnamefont {Ritsch}},\ }\bibfield  {title} {\bibinfo {title} {Quantumoptics.jl: A {Julia} framework for simulating open quantum systems},\ }\href {https://doi.org/10.1016/j.cpc.2018.02.004} {\bibfield  {journal} {\bibinfo  {journal} {Comput. Phys. Commun.}\ }\textbf {\bibinfo {volume} {227}},\ \bibinfo {pages} {109} (\bibinfo {year} {2018})}\BibitemShut {NoStop}%
\end{thebibliography}
\end{document}